\documentclass[twocolumn,amsmath,amssymb,superscriptaddress]{revtex4-2}

\usepackage{graphicx}
\usepackage{hyperref}
\usepackage{braket}
\usepackage{bbm}

\DeclareMathOperator{\tr}{Tr}
\DeclareMathOperator{\imag}{Im}
\DeclareMathOperator{\var}{Var}

\newcommand{\avg}[1]{\langle #1 \rangle}

\usepackage[svgnames]{xcolor}
\usepackage{ulem}
\begin{document}

\title{Invariance property of the Fisher information in scattering media}

\author{Michael Horodynski}
\affiliation{Institute for Theoretical Physics, Vienna University of Technology (TU Wien), Vienna, Austria}
\author{Dorian Bouchet}
\affiliation{Université Grenoble Alpes, CNRS, LIPhy, Grenoble, France}
\author{Stefan Rotter}
\email{stefan.rotter@tuwien.ac.at}
\affiliation{Institute for Theoretical Physics, Vienna University of Technology (TU Wien), Vienna, Austria}

\begin{abstract}
    Determining the ultimate precision limit for measurements on a sub-wavelength particle with coherent laser light is a goal with applications in areas as diverse as biophysics and nanotechnology. Here, we demonstrate that surrounding such a particle with a complex scattering environment does, on average, not have any influence on the mean quantum Fisher information associated with measurements on the particle. As a remarkable consequence, the average precision that can be achieved when estimating the particle's properties is the same in the ballistic and in the diffusive scattering regime, independently of the particle's position within its complex environment. This invariance law breaks down only in the regime of Anderson localization, due to increased $C_0$–speckle correlations. Finally, we show how these results connect to the mean quantum Fisher information achievable with spatially optimized input fields.
\end{abstract}

\maketitle

Precisely estimating the properties of sub-wavelength particles surrounded by scattering environments is a central aspect of many research areas, ranging from the measurement of the position and the mass of biological molecules~\cite{taylor_interferometric_2019,young_interferometric_2019} to the characterization of engineered nanostructured samples~\cite{orji_metrology_2018}. In these contexts, multiple scattering effects are usually seen as a major drawback, limiting the achievable precision in the estimation of observable parameters characterizing the particles. Nevertheless, it is also known that such multiple scattering effects can sometimes be beneficial to optical imaging~\cite{simonetti_multiple_2006,girard_nanometric_2010,PhysRevLett.116.073902,PhysRevResearch.2.033148} as well as to single-particle localization and sensing~\cite{PhysRevLett.65.3120,PhysRevB.43.8638,denOuter:93,berk2021multiple}, especially if prior knowledge on the scattering environment is available to the observer~\cite{szameit_sparsity-based_2012,zhang_far-field_2016}. These insights naturally raise the question of how the presence of such a complex environment surrounding a sub-wavelength particle influences the information carried by the field scattered by the particle.

In recent years, the concept of Fisher information has enabled considerable progress in the precise characterization of sub-wavelength particles through optical measurements. Notably, this concept was used to analytically derive fundamental limits on the achievable localization precision in the case of a fluorescent molecule located in free space~\cite{ober_localization_2004,mortensen_optimized_2010,deschout_precisely_2014,backlund_fundamental_2018} and for estimating the distance between two incoherent point sources~\cite{ram_beyond_2006,tsang_quantum_2016,paur_achieving_2016,Zhou:19}. In parallel, the influence of multiple scattering effects upon the Fisher information has been investigated based on numerical approaches for two spherical particles~\cite{sentenac_influence_2007} as well as for larger ensembles of sub-wavelength particles~\cite{bouchet_influence_2020}. These studies show that multiple scattering effects can either increase or decrease the achievable localization precision, depending on the microstructure of the scattering environment. 
\begin{figure}[t!]
    \centering
    \includegraphics[width=\columnwidth]{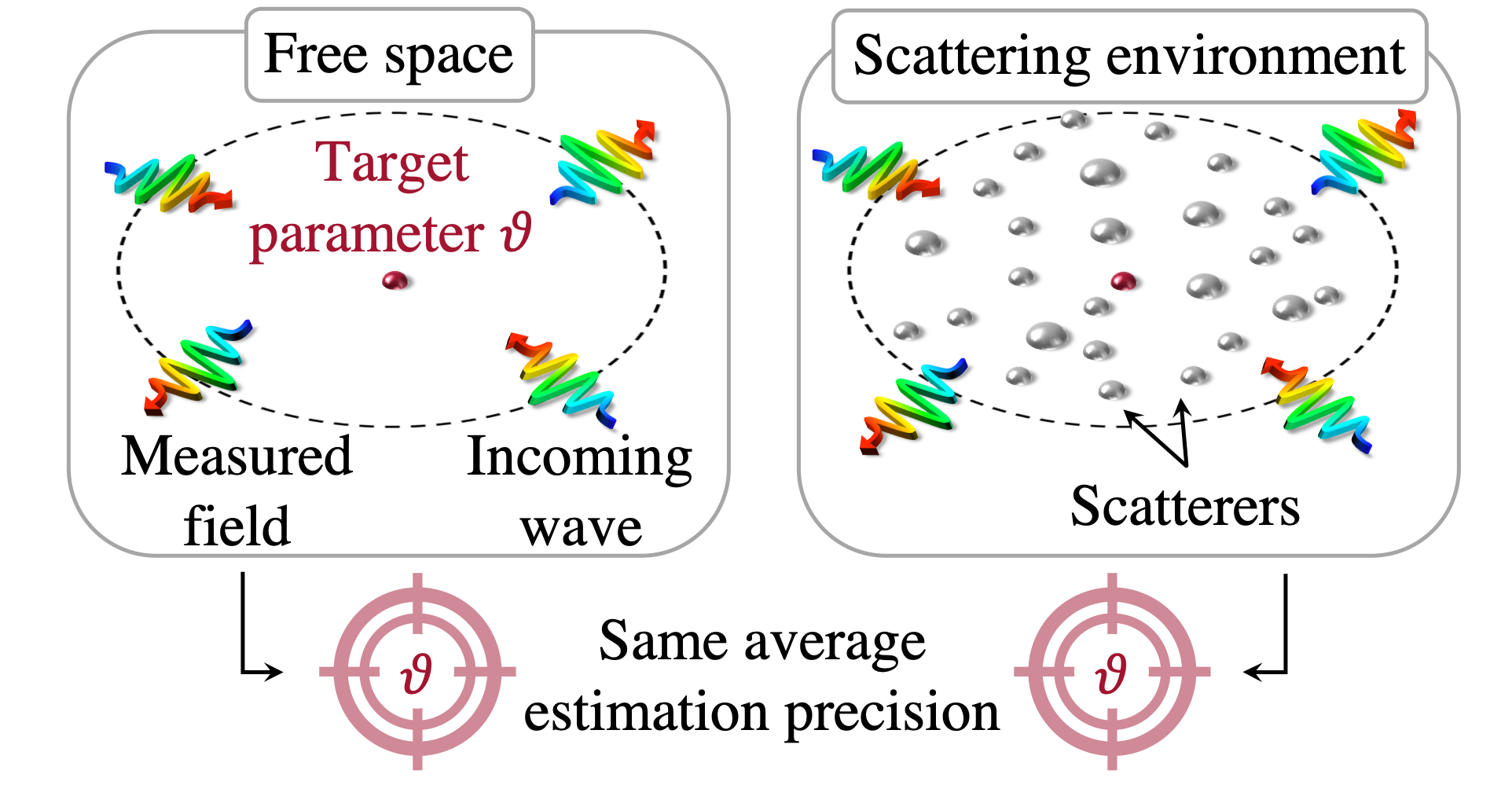}
    \caption{Illustration of the concept. When illuminating a target (red) with coherent light (incoming arrows), the average precision that can be achieved when estimating a property $\theta$ of the target is the same whether the target is freely accessible (left) or embedded inside a scattering environment (right).}
    \label{fig:cf}
\end{figure}

In this Letter, we uncover that the measurement precision achievable in scattering systems is actually subject to a counter-intuitive invariance rule: the scattering environment surrounding a sub-wavelength particle does, on average, \emph{not} influence \emph{at all} the precision that can be achieved when estimating the properties of this particle (see Fig.\ \ref{fig:cf}). This result sheds new light on the possibility to use multiple-scattering effects for improving the measurement precision in complex systems. Moreover, the intimate connection between measurement and back-action \cite{bouchet_maximum_2021} allows us to interpret this result also as an invariance rule for the average micro-manipulation capabilites of waves in a scattering environment. 

Specifically, we show that, in the canonical case of a flux-conserving system illuminated by a coherent field, the quantum Fisher information (QFI)---averaged over input angles, frequency and configurations---as well as the average quantum Cramér-Rao bound (QCRB) are both independent of the scattering strength of the environment surrounding the particle and on the particle's position. This invariance law holds both in the ballistic and in the diffusive regime, whereas deviations are observed when Anderson localization sets in. Remarkably, in this case, both the average QFI and the average QCRB increase with the scattering strength of the medium as a result of $C_0$–speckle correlations. In addition, we quantify the increase of the average QFI obtained when the incident field is spatially optimized using wavefront shaping techniques. Originating from a fundamental connection between the QFI~\cite{helstrom_quantum_1969}, the local density of states (LDOS)~\cite{barnes_fluorescence_1998} and the cross density of states (CDOS)~\cite{caze_spatial_2013}, these results show that a simple fundamental law rules the ultimate precision limit achievable in estimating properties of a sub-wavelength particles using coherent light scattering.

The unavoidable presence of measurement noise imposes a lower limit on the precision that can be reached when using scattering measurements to estimate any given observable parameter $\theta$, such as the position or the dielectric constant of a sub-wavelength particle. The ultimate precision limit, as governed by the quantum fluctuations of the probe field, is then determined by the QFI defined as $\mathcal{I}_\theta=\tr (\rho L^2)$, where the density operator $\rho$ describes the quantum state of the light and $L$ denotes the symmetrized logarithmic derivative \cite{helstrom_quantum_1969,braunstein_generalized_1996}. The QCRB, which bounds the variance of unbiased estimators of $\theta$, is simply expressed by $\Sigma_\theta=1/\mathcal{I}_\theta$. This bound is reachable using an optimal detection scheme \cite{braunstein_generalized_1996} together with an efficient estimator, which is easily found in the limit of small parameter variations~\cite{trees_detection_2013}. The QFI thus provides us with a relevant metric to compare the estimation precision achievable with different probe fields and different scattering environments.

Assuming that the probe field is in a coherent state and that the scattering matrix $S$ of the system is known, it can be shown that the QFI takes on the following quadratic form~\cite{bouchet_maximum_2021},
\begin{align}\label{eq:eq1}
    \mathcal{I}_\theta = 4 \braket{u | F_\theta | u}\,,
\end{align}
where $\ket{u}$ is an arbitrary asymptotic wave state impinging on the system (typically a vector of modal amplitudes), $F_\theta=\partial_\theta S^\dagger \partial_\theta S$ is the so-called Fisher information operator for the estimation of $\theta$ and $\dagger$ denotes the conjugate transpose. For any arbitrary incident wave state, Eq.~\eqref{eq:eq1} can thus be used to predict the QFI relative to the estimation of the parameter $\theta$. Moreover, in the ideal case of a unitary $S$-matrix, we can write $F_\theta = Q_\theta^2$, where $Q_\theta = -\mathrm{i} S^{-1} \partial_\theta S$ denotes the Generalized Wigner-Smith (GWS) operator~\cite{ambichl_focusing_2017,horodynski_optimal_2020}. Due to this connection, Eq.~\eqref{eq:eq1} can equivalently be used to quantify the perturbation applied to the parameter $\theta$ by the probe field. In the following, we will assume that the field is sufficiently weak, so that it does not significantly perturb the value of $\theta$ during the measurement process.
\begin{figure*}[t!]
	\centering
	\includegraphics[width=\textwidth]{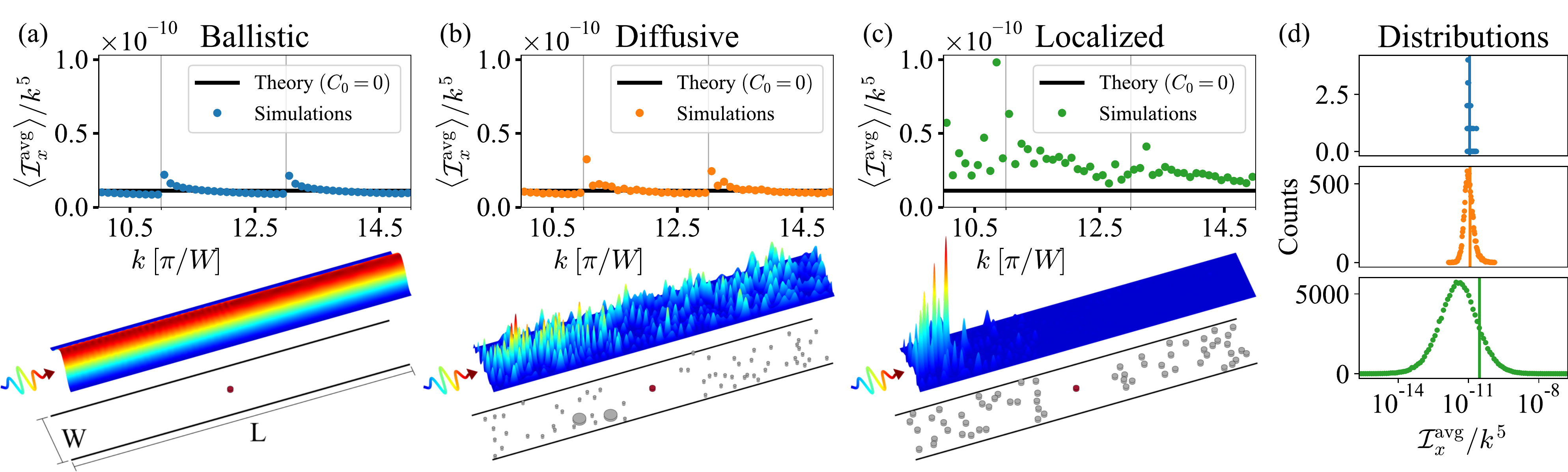}
	\caption{Average QFI relative to the position $x$ of a sub-wavelength particle in (a) the ballistic, (b) the diffusive, and (c) the localized regime. In the ballistic and diffusive regimes, numerical results are in excellent agreement with the theoretical predictions of Eq.~\eqref{eq:eq4}, calculated for $C_0=0$ (black lines). In the localized regime, the $C_0$–speckle correlations cannot be neglected, leading to an increase of the average QFI. In the diffusive and localized regimes, averages are calculated from 250 and 2500 different random configurations, respectively. Note that we approximated the number of transverse waveguide modes $N \simeq 2kW/\pi$ (grey vertical bars indicate odd mode openings). 
	(a-c, lower panels) The intensity distribution of the fundamental mode injected from the left lead is depicted (a) for an empty waveguide and for a given configuration of scatterers (grey cylinders) (b) with refractive index $n=1.44$ or (c) with hard walls. The target (red cylinder) has refractive index $n=1.44$ and is not shown to scale. As compared to the length $L$ of the waveguide, the transport mean free path is $\ell_\mathrm{tr} \approx  0.47L$ in the diffusive regime and the localization length is $\xi\approx 0.4L$ in the localized regime. (d) Distribution (on a logarithmic scale) of the average QFI $\mathcal{I}^\mathrm{avg}_\theta$ for different disorder realizations, in the ballistic (blue), diffusive (orange) and localized (green) regimes, together with the value of $ \langle \mathcal{I}^\mathrm{avg}_\theta \rangle $ (vertical lines), which is averaged over disorder realizations.}
	\label{fig:fig1}
\end{figure*}

The QFI averaged over all possible input fields is given by the following trace:
\begin{align}\label{eq:eq2}
    \mathcal{I}^\mathrm{avg}_\theta  = \frac{4 }{N} \tr F_\theta = \frac{16}{N\Delta \theta^2} \tr \left[\imag \left(G\right) \Delta H\right]^2,
\end{align}
where $N$ is the total number of flux-carrying incoming channels. We arrive at the second equality of Eq.~\eqref{eq:eq2} by using the expansion of the GWS operator $Q_\theta = 2 V^\dagger G^\dagger \Delta H G V/ \Delta \theta$, where $G$ is the Green's function, $V$ is the coupling matrix of the scattering system to the asymptotic channels,  $\Delta \theta$ is an infinitesimal variation of the parameter of interest, and $\Delta H$ is the corresponding perturbation in the system Hamiltonian~\cite{horodynski_optimal_2020}. Equation~\eqref{eq:eq2} expresses a system-specific connection between the externally-accessible quantity $\mathcal{I}^\mathrm{avg}_\theta$ and the local quantity $\imag G$, which is proportional to the LDOS and CDOS at the target position. We will now show how this connection turns out to be extremely useful to demonstrate that the information is (on average) independent of system-specific parameters.

To this end, we now focus our attention on the estimation of a parameter $\theta$ characterizing a sub-wavelength particle, such as one of its coordinates (denoted by $x$) or its dielectric constant (denoted by $\varepsilon$). We also work under the assumption that the spatial distribution of the light field is statistically homogeneous and isotropic throughout the scattering medium. This occurs when the scattering medium itself as well as the incoming light are statistically homogeneous and isotropic (i.e., for Lambertian illumination) \cite{savo_observation_2017,benichou_averaged_2005}. We are then allowed to use the concept of the CDOS in order to quantify the effect of a small change in the position of the target on the QFI~\cite{caze_spatial_2013,canaguier-durand_cross_2019}, resulting in a simple relation between $\mathcal{I}^\mathrm{avg}_\theta$ and the LDOS $\rho(\mathbf{r}_T,k)$ at the location $\mathbf{r}_T$ of the target (see supplementary material). As a result, we find that the QFI averaged over all input channels $(\mathrm{avg})$
and disorder configurations ($\langle\ldots\rangle$) reads
\begin{equation}\label{eq:eq3}
    \avg{\mathcal{I}^\mathrm{avg}_\theta}   = \frac{4 B_\theta k^2\pi^2}{N \varepsilon^2} \avg{\rho^2(\mathbf{r}_T,k)} ,
\end{equation}
where the prefactor $B_\theta$ depends on the parameter of interest such that $B_x = k^2 (\varepsilon-1)^2$ if $\theta=x$ and $B_\varepsilon = 1$ if $\theta=\varepsilon$. Importantly, Eq.~\eqref{eq:eq3} links the average QFI directly to the square of the LDOS, thereby establishing that properties of particles that are placed at positions with an increased LDOS can be estimated more accurately. Our derivation also shows that a linear scaling of the LDOS with the QFI occurs in the case of considerably sub-unitary scattering matrices (see the supplementary material for a detailed derivation), as numerically observed in Ref.~\cite{bouchet_influence_2020}.

Under the isotropy assumption, we can use Weyl's law---which gives a simple expression for the average DOS in the asymptotic limit \cite{weyl_ueber_1911,arendt_mathematical_2009}---to estimate the average LDOS, $\langle \rho(\mathbf{r}_T,k) \rangle = Ak\varepsilon/(2\pi)$ (in 2D). To use this approximation, however, we first have to relate $\avg{\rho^2(\mathbf{r}_T,k)}$ to $\avg{\rho(\mathbf{r}_T,k)}^2$. For this purpose, we use the connection between the variance of the LDOS and the $C_0$–speckle correlations,  $C_0 = \var [\rho(\mathbf{r}_T,k)]/\avg{ \rho(\mathbf{r}_T,k) }^2$ \cite{van_tiggelen_fluctuations_2006,caze_near-field_2010}, resulting in: 
\begin{equation}\label{eq:eq4}
    \avg{\mathcal{I}^\mathrm{avg}_\theta} =  \frac{B_\theta k^4 A^2}{N}  \left( 1+C_0 \right)\,.
\end{equation}
The above equation establishes a counter-intuitive invariance law: the average QFI is independent of the target's surrounding environment and of its position therein–––provided that light scattering is in the ballistic or diffusive regime (for which $C_0 \simeq 0$, as numerically confirmed in the supplementary material). Due to the connection between information and back-action, this result also implies that the average magnitude of the momentum transferred onto the particle as well as the average intensity focused onto it are invariant quantities with respect to the scattering strength. Note that, while this invariance law is derived here for the 2-dimensional scalar Helmholtz equation, the generality of the Fisher information operator and of the LDOS suggests that similar results could also be derived with a 3-dimensional, vectorial model.

Equation~\eqref{eq:eq3} also reveals an interesting conceptual connection to another known invariant quantity in wave scattering: the mean path length, which is related to the average density of states (DOS) in the whole scattering volume~\cite{pierrat_invariance_2014,savo_observation_2017,davy_mean_2021}. In a similar manner, we show here how the invariance of the mean path length in every sub-volume \cite{benichou_averaged_2005} and an equivalent invariance of the LDOS are connected to the Fisher information.

To illustrate the implications of this invariance law, we choose the case of position estimations ($\theta=x$) for which we perform numerical simulations using a finite-element method (NGSolve) \cite{schoberl_netgen_1997,schoberl_c11_2014}. Our model system consists of a rectangular multimode waveguide of width $W$ and length $L$, with hard walls at the top and bottom, and with openings to the left and right (see lower panels of Fig.~\ref{fig:fig1}). In this waveguide, we randomly place scatterers of different sizes and scattering cross-sections in order to simulate disorder of adjustable scattering strength. We vary the wavelength between $\lambda = 2\pi/k \approx 0.2W$ and $\lambda \approx 0.13W$ such that between $10$ and $14$ transverse modes propagate inside the waveguide. The radius of the circular target is set to $R=W/200$, which is an order of magnitude smaller than the minimal wavelength considered here. For each configuration of the scatterers, after numerically calculating the scattering matrices associated with two close-by positions of the target, the derivative of scattering matrices with respect to the coordinate $x$ is determined with a finite-difference scheme and the associated QFI is obtained using Eq.~\eqref{eq:eq1}.

Using this numerical framework, we separately investigate the ballistic, diffusive and localized regimes. For both ballistic and diffusive scattering (see Fig.~\ref{fig:fig1}a,b), our numerical results agree very well with the theoretical prediction of Eq.~\eqref{eq:eq4} when the $C_0$–speckle correlation is neglected ($C_0\simeq 0$). We note here that near-field interactions between scatterers can change the behavior of the $C_0$–speckle correlations \cite{sapienza_long-tail_2011}—an effect we have explicitly excluded by not placing any scatterers within an exclusion square of sidelength $W$ around the target. These results thus confirm that, even though the QFI strongly varies from configuration to configuration (see Fig.~\ref{fig:fig1}d), its average value stays remarkably insensitive to the scattering strength of the environment. Nevertheless, in the localized regime (Fig.~\ref{fig:fig1}c), $C_0$ cannot be neglected anymore and the average QFI increases. This trend can be explained by the estimate $C_0 \approx \pi/k\ell_\mathrm{tr}$ \cite{van_tiggelen_fluctuations_2006}, where $\ell_\mathrm{tr}$ is the transport mean free path (see also the supplementary material for a numerical estimate of the $C_0$ correlation function). This increase of the average QFI in the localized regime can be understood from the fact that, while most configurations are associated with a very low QFI, a few configurations display an exceptionally large QFI (see Fig.~\ref{fig:fig1}d), resulting overall in an increase of its average value.

\begin{figure}[t!]
	\centering
	\includegraphics[width=\columnwidth]{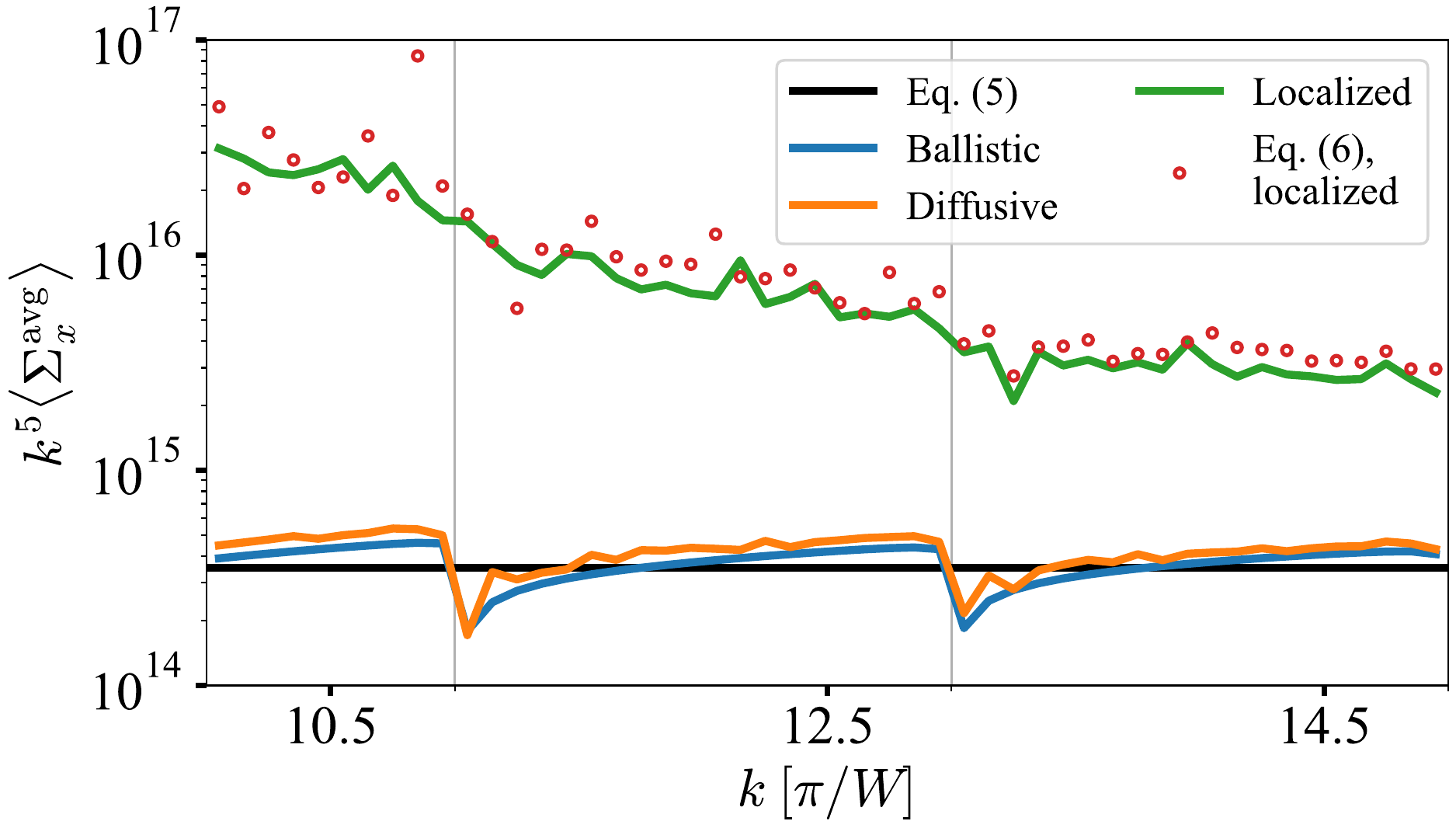}
	\caption{Average QCRB $\avg{\Sigma_\theta^{\mathrm{avg}}}$ relative to the position $x$ for a subwavelength particle in the ballistic (blue), diffusive (orange) and localized (green) regime (grey vertical bars indicate odd mode openings in the waveguide). In the ballistic and diffusive regimes, that are most easily accessible in optics, numerical results (blue and orange lines) are in excellent agreement with the theoretical predictions of Eq.~\eqref{eq:eq6}, demonstrating that the invariance law of the QFI leads to a similar invariance for the QCRB in these regimes. By contrast, in the localized regime, the average QCRB is drastically increased. In this regime, where linear propagation of error cannot be used, the average QCRB is correctly predicted using Eq.~\eqref{eq:eq5} (red dots).}
	\label{fig:fig4}
\end{figure}

The average precision that can be achieved in an experiment is obtained by calculating $\avg{\Sigma_\theta^{\mathrm{avg}}}$, which is the QCRB averaged over input fields and disorder configurations. This quantity is shown in Fig.~\ref{fig:fig4} for the three scattering regimes previously studied (ballistic, diffusive and localized). In the ballistic and diffusive regimes, that are easily accessible in optics, we observe that the average QCRB is constant regardless of the scattering strength of the environment, in the same way as the average QFI. Indeed, $\avg{\Sigma_\theta^{\mathrm{avg}}}$ is here correctly estimated from linear propagation of error, i.e., $\avg{\Sigma_\theta^{\mathrm{avg}}} \simeq 1/\avg{\mathcal{I}_\theta^{\mathrm{avg}}}$. Since $C_0\simeq0$ in these regimes, we obtain
\begin{equation}\label{eq:eq6}
\avg{\Sigma_\theta^{\mathrm{avg}}} \simeq \frac{N}{B_\theta k^4 A^2} .
\end{equation}
This expression shows that a simple invariance law also rules the average precision that can be achieved when estimating the properties of a sub-wavelength particle. This law yields the remarkable insight that the average precision achievable for a particle in free space neither increases nor decreases when this particle is placed in a complex scattering environment.

Nevertheless, this invariance law does not apply to the localized regime, in which the average QCRB drastically increases with the scattering strength of the environment (see Fig.~\ref{fig:fig4}c). In this regime, linear propagation of error cannot be used, and $\avg{\Sigma_\theta^{\mathrm{avg}}} \neq 1/\avg{\mathcal{I}_\theta^{\mathrm{avg}}}$. A connection can, however, still be made using the observation that both quantities follow a log-normal distribution \cite{bouchet_influence_2020} (see the supplementary material for a detailed derivation), resulting in:
\begin{equation}\label{eq:eq5}
    \avg{\Sigma_\theta^{\mathrm{avg}}} = \frac{\avg{\tr ^2 F_\theta}}{4N \avg{\tr F_\theta}^3} .
\end{equation}
This expression not only involves the variance of the LDOS but also higher moments of the LDOS and the CDOS (see supplementary material), which can be numerically estimated with a maximum likelihood method. The theoretical prediction obtained using this procedure are in excellent agreement with numerical observations, as shown in Fig.~\ref{fig:fig4}.

Routine access to light's spatial degrees of freedom \cite{mosk_controlling_2012,rotter_light_2017} enables the generation of input fields that are spatially optimized to maximize the Fisher information \cite{bouchet_maximum_2021}. This naturally leads to the question whether such a simple law as Eq.~\eqref{eq:eq4} also exists for such spatially-optimized input fields. In general, the QFI, first maximized over input fields and then averaged over the maxima of many disorder configurations, can be written as
\begin{equation}\label{eq:eq7}
    \langle \mathcal{I}^\mathrm{max}_\theta \rangle = p_\theta N \langle \mathcal{I}_\theta^\mathrm{avg} \rangle,
\end{equation}
where $p_\theta=\langle \Lambda_1 \rangle/\langle \tr F_\theta \rangle$, with $\Lambda_1$ being the largest eigenvalue of $F_\theta$. Note that Eq.~\eqref{eq:eq7} is a simple reformulation of the identity $\avg{\mathcal{I}^\mathrm{max}_\theta} = 4 \avg{\Lambda_1}$, which follows directly from Eq.~\eqref{eq:eq1}. In order to assess the benefits of maximizing the QFI over input fields, we need to determine the value of $p_\theta$, which depends upon the parameter of interest.

\begin{figure}[t!]
	\centering
	\includegraphics[width=\columnwidth]{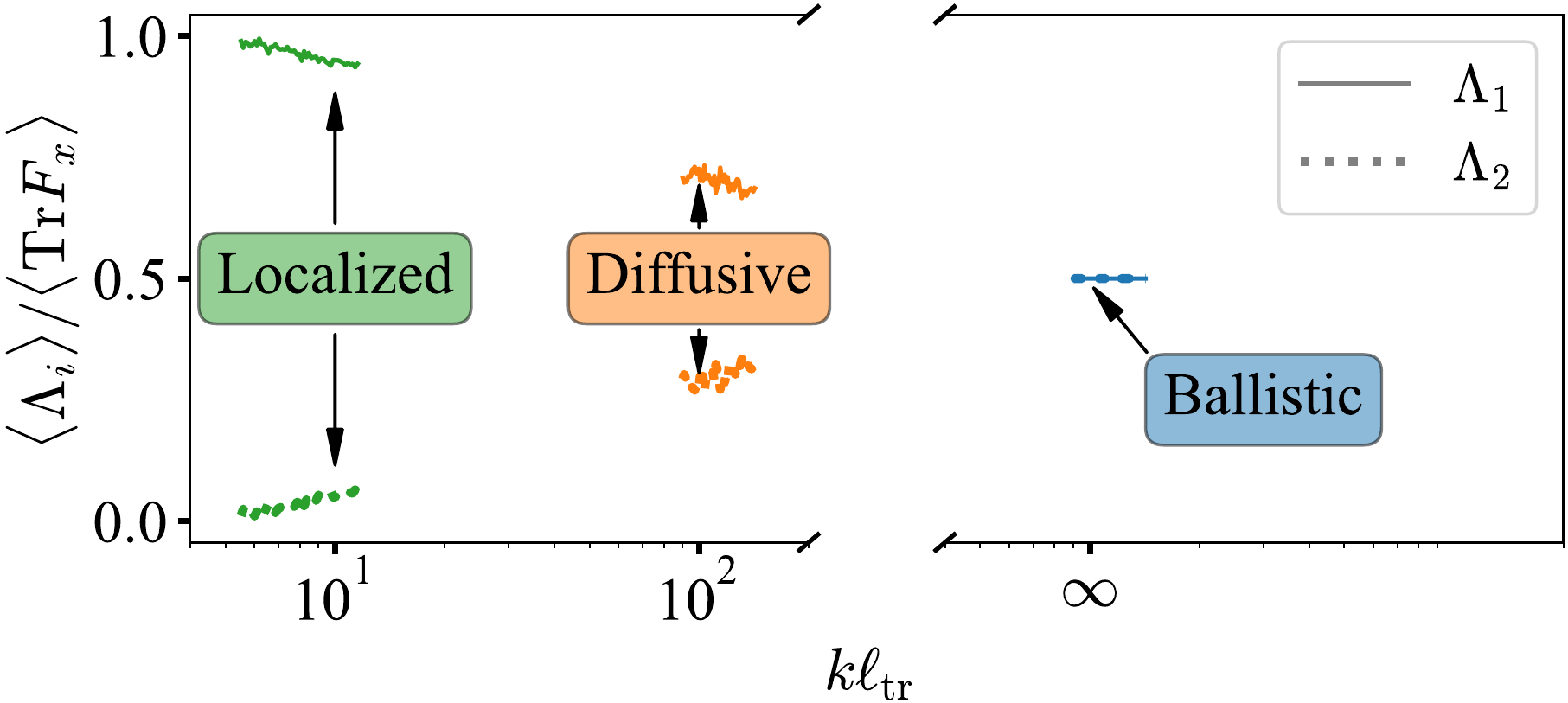}
	\caption{Average normalized magnitude of the two largest eigenvalues, $\Lambda_1, \Lambda_2$, of $F_x$ as a function of $k\ell_\mathrm{tr}$. In the ballistic regime, the rank of $F_x$ is equal to 2, with equal contributions from the two non-zero eigenvalues ($\Lambda_1/\Lambda_2 \approx 1$). This symmetry breaks with increased scattering strength, resulting in an increased contribution from the largest eigenvalue in the diffusive and localized regimes. The numerical data used in this figure also underlies Figs.~\ref{fig:fig1}~and~\ref{fig:fig4}.}
	\label{fig:fig3}
\end{figure}

On the one hand, for dielectric-constant estimations of a point-like target, we can write the following identity: $\tr F_\varepsilon = \tr Q_\varepsilon^2 =  \tr^2 Q_\varepsilon$. This implies that the rank of $F_\varepsilon$ is $1$ regardless of the scattering environment, resulting in $p_\varepsilon=1$. On the other hand, concerning estimations of the target's position, we can assess the value of $p_x$ by studying the momentum delivered on the target by the probe field. Since, in the ballistic case, this momentum transfer can be applied both in the positive and negative $x$-direction, we can expect a value of $2$ for the rank of $F_x$, resulting in $p_x=0.5$. To confirm this reasoning, we use numerical simulations to calculate the ratios $\Lambda_1/\tr F_x$ and $\Lambda_2/\tr F_x$ (involving the largest and second largest eigenvalue $\Lambda_{1,2}$ of $F_\theta$) for different disorder configurations and different scattering strength. In Fig.~\ref{fig:fig3}, we can see that these two quantities are, indeed, equal to $0.5$ in the ballistic case, resulting in $p_x=0.5$. However, adding a scattering environment around the particle breaks the directional symmetry, such that the normalized magnitudes of the two non-zero eigenvalues begin to bifurcate. An increase of the scattering strength leads to a stronger symmetry breaking; $p_x$ thus increases with the scattering strength until it approaches its maximal value of $p_x=1$ in the localized regime. The physical intuition emerging from these results is that the increase in QFI in the localized regime comes at the prize of reduced micro-manipulation capabilities (only one direction can be controlled).

To summarize, we derive a universal invariance property for the average QFI and for the average QCRB applicable to both the ballistic and diffusive regimes. This result implies that coherent light fields can be used to measure the position or the dielectric constant of a sub-wavelength particle with an average precision that is the same, regardless of the scattering strength of the surrounding environment. In the regime of Anderson localization, however, the increased scattering strength simultaneously enhances the average QFI and the average QCRB. Since strong interaction and precise measurements are two sides of the same coin \cite{bouchet_maximum_2021}, the derived invariance property equivalently applies to the average magnitude of the momentum transferred by the probe field to a sub-wavelength particle. The maximum applicable force, in turn, grows continuously with increasing scattering strength even in the localized regime, at the expense, however, of a decreasing control over the force direction. Although these results are derived here for a 2-dimensional model system in the present work, extensions to all physical scenarios featuring a linear wave equation should be possible. In particular, our results should be extendable to targets in nonuniform engineered media and correlated disorders, such as photonic crystals~\cite{koenderink_optical_2005}, Levy glasses~\cite{barthelemy_levy_2008} and hyperuniform media~\cite{torquato_local_2003} as well as to inhomogeneously disordered materials \cite{huang_invariance_2020}. Our derivation also paves the ground for the analysis of lossy scattering media, such as biological tissue, and for the characterization of extended targets. To this end, a deeper exploration of the fundamental connection between the LDOS, the CDOS and the QFI would certainly be of considerable interest.

Support by the Austrian Science Fund (FWF) under Project No.\ P32300 (WAVELAND) is gratefully acknowledged. The computational results were achieved using the Vienna Scientific Cluster (VSC).

\bibliographystyle{apsrev4-2}

\begin{thebibliography}{49}%
\makeatletter
\providecommand \@ifxundefined [1]{%
 \@ifx{#1\undefined}
}%
\providecommand \@ifnum [1]{%
 \ifnum #1\expandafter \@firstoftwo
 \else \expandafter \@secondoftwo
 \fi
}%
\providecommand \@ifx [1]{%
 \ifx #1\expandafter \@firstoftwo
 \else \expandafter \@secondoftwo
 \fi
}%
\providecommand \natexlab [1]{#1}%
\providecommand \enquote  [1]{``#1''}%
\providecommand \bibnamefont  [1]{#1}%
\providecommand \bibfnamefont [1]{#1}%
\providecommand \citenamefont [1]{#1}%
\providecommand \href@noop [0]{\@secondoftwo}%
\providecommand \href [0]{\begingroup \@sanitize@url \@href}%
\providecommand \@href[1]{\@@startlink{#1}\@@href}%
\providecommand \@@href[1]{\endgroup#1\@@endlink}%
\providecommand \@sanitize@url [0]{\catcode `\\12\catcode `\$12\catcode
  `\&12\catcode `\#12\catcode `\^12\catcode `\_12\catcode `\%12\relax}%
\providecommand \@@startlink[1]{}%
\providecommand \@@endlink[0]{}%
\providecommand \url  [0]{\begingroup\@sanitize@url \@url }%
\providecommand \@url [1]{\endgroup\@href {#1}{\urlprefix }}%
\providecommand \urlprefix  [0]{URL }%
\providecommand \Eprint [0]{\href }%
\providecommand \doibase [0]{https://doi.org/}%
\providecommand \selectlanguage [0]{\@gobble}%
\providecommand \bibinfo  [0]{\@secondoftwo}%
\providecommand \bibfield  [0]{\@secondoftwo}%
\providecommand \translation [1]{[#1]}%
\providecommand \BibitemOpen [0]{}%
\providecommand \bibitemStop [0]{}%
\providecommand \bibitemNoStop [0]{.\EOS\space}%
\providecommand \EOS [0]{\spacefactor3000\relax}%
\providecommand \BibitemShut  [1]{\csname bibitem#1\endcsname}%
\let\auto@bib@innerbib\@empty
\bibitem [{\citenamefont {Taylor}\ and\ \citenamefont
  {Sandoghdar}(2019)}]{taylor_interferometric_2019}%
  \BibitemOpen
  \bibfield  {author} {\bibinfo {author} {\bibfnamefont {R.~W.}\ \bibnamefont
  {Taylor}}\ and\ \bibinfo {author} {\bibfnamefont {V.}~\bibnamefont
  {Sandoghdar}},\ }\href {https://doi.org/10.1021/acs.nanolett.9b01822}
  {\bibfield  {journal} {\bibinfo  {journal} {Nano Letters}\ }\textbf {\bibinfo
  {volume} {19}},\ \bibinfo {pages} {4827} (\bibinfo {year}
  {2019})}\BibitemShut {NoStop}%
\bibitem [{\citenamefont {Young}\ and\ \citenamefont
  {Kukura}(2019)}]{young_interferometric_2019}%
  \BibitemOpen
  \bibfield  {author} {\bibinfo {author} {\bibfnamefont {G.}~\bibnamefont
  {Young}}\ and\ \bibinfo {author} {\bibfnamefont {P.}~\bibnamefont {Kukura}},\
  }\href {https://doi.org/10.1146/annurev-physchem-050317-021247} {\bibfield
  {journal} {\bibinfo  {journal} {Annu. Rev. Phys. Chem.}\ }\textbf {\bibinfo
  {volume} {70}},\ \bibinfo {pages} {301} (\bibinfo {year} {2019})}\BibitemShut
  {NoStop}%
\bibitem [{\citenamefont {Orji}\ \emph {et~al.}(2018)\citenamefont {Orji},
  \citenamefont {Badaroglu}, \citenamefont {Barnes}, \citenamefont {Beitia},
  \citenamefont {Bunday}, \citenamefont {Celano}, \citenamefont {Kline},
  \citenamefont {Neisser}, \citenamefont {Obeng},\ and\ \citenamefont
  {Vladar}}]{orji_metrology_2018}%
  \BibitemOpen
  \bibfield  {author} {\bibinfo {author} {\bibfnamefont {N.~G.}\ \bibnamefont
  {Orji}}, \bibinfo {author} {\bibfnamefont {M.}~\bibnamefont {Badaroglu}},
  \bibinfo {author} {\bibfnamefont {B.~M.}\ \bibnamefont {Barnes}}, \bibinfo
  {author} {\bibfnamefont {C.}~\bibnamefont {Beitia}}, \bibinfo {author}
  {\bibfnamefont {B.~D.}\ \bibnamefont {Bunday}}, \bibinfo {author}
  {\bibfnamefont {U.}~\bibnamefont {Celano}}, \bibinfo {author} {\bibfnamefont
  {R.~J.}\ \bibnamefont {Kline}}, \bibinfo {author} {\bibfnamefont
  {M.}~\bibnamefont {Neisser}}, \bibinfo {author} {\bibfnamefont
  {Y.}~\bibnamefont {Obeng}},\ and\ \bibinfo {author} {\bibfnamefont {A.~E.}\
  \bibnamefont {Vladar}},\ }\href {https://doi.org/10.1038/s41928-018-0150-9}
  {\bibfield  {journal} {\bibinfo  {journal} {Nature Electronics}\ }\textbf
  {\bibinfo {volume} {1}},\ \bibinfo {pages} {532} (\bibinfo {year}
  {2018})}\BibitemShut {NoStop}%
\bibitem [{\citenamefont {Simonetti}(2006)}]{simonetti_multiple_2006}%
  \BibitemOpen
  \bibfield  {author} {\bibinfo {author} {\bibfnamefont {F.}~\bibnamefont
  {Simonetti}},\ }\href {https://doi.org/10.1103/PhysRevE.73.036619} {\bibfield
   {journal} {\bibinfo  {journal} {Phys. Rev. E}\ }\textbf {\bibinfo {volume}
  {73}},\ \bibinfo {pages} {036619} (\bibinfo {year} {2006})}\BibitemShut
  {NoStop}%
\bibitem [{\citenamefont {Girard}\ \emph {et~al.}(2010)\citenamefont {Girard},
  \citenamefont {Maire}, \citenamefont {Giovannini}, \citenamefont {Talneau},
  \citenamefont {Belkebir}, \citenamefont {Chaumet},\ and\ \citenamefont
  {Sentenac}}]{girard_nanometric_2010}%
  \BibitemOpen
  \bibfield  {author} {\bibinfo {author} {\bibfnamefont {J.}~\bibnamefont
  {Girard}}, \bibinfo {author} {\bibfnamefont {G.}~\bibnamefont {Maire}},
  \bibinfo {author} {\bibfnamefont {H.}~\bibnamefont {Giovannini}}, \bibinfo
  {author} {\bibfnamefont {A.}~\bibnamefont {Talneau}}, \bibinfo {author}
  {\bibfnamefont {K.}~\bibnamefont {Belkebir}}, \bibinfo {author}
  {\bibfnamefont {P.~C.}\ \bibnamefont {Chaumet}},\ and\ \bibinfo {author}
  {\bibfnamefont {A.}~\bibnamefont {Sentenac}},\ }\href
  {https://doi.org/10.1103/PhysRevA.82.061801} {\bibfield  {journal} {\bibinfo
  {journal} {Phys. Rev. A}\ }\textbf {\bibinfo {volume} {82}},\ \bibinfo
  {pages} {061801(R)} (\bibinfo {year} {2010})}\BibitemShut {NoStop}%
\bibitem [{\citenamefont {Newman}\ \emph {et~al.}(2016)\citenamefont {Newman},
  \citenamefont {Luo},\ and\ \citenamefont {Webb}}]{PhysRevLett.116.073902}%
  \BibitemOpen
  \bibfield  {author} {\bibinfo {author} {\bibfnamefont {J.~A.}\ \bibnamefont
  {Newman}}, \bibinfo {author} {\bibfnamefont {Q.}~\bibnamefont {Luo}},\ and\
  \bibinfo {author} {\bibfnamefont {K.~J.}\ \bibnamefont {Webb}},\ }\href
  {https://doi.org/10.1103/PhysRevLett.116.073902} {\bibfield  {journal}
  {\bibinfo  {journal} {Phys. Rev. Lett.}\ }\textbf {\bibinfo {volume} {116}},\
  \bibinfo {pages} {073902} (\bibinfo {year} {2016})}\BibitemShut {NoStop}%
\bibitem [{\citenamefont {Luo}\ and\ \citenamefont
  {Webb}(2020)}]{PhysRevResearch.2.033148}%
  \BibitemOpen
  \bibfield  {author} {\bibinfo {author} {\bibfnamefont {Q.}~\bibnamefont
  {Luo}}\ and\ \bibinfo {author} {\bibfnamefont {K.~J.}\ \bibnamefont {Webb}},\
  }\href {https://doi.org/10.1103/PhysRevResearch.2.033148} {\bibfield
  {journal} {\bibinfo  {journal} {Phys. Rev. Research}\ }\textbf {\bibinfo
  {volume} {2}},\ \bibinfo {pages} {033148} (\bibinfo {year}
  {2020})}\BibitemShut {NoStop}%
\bibitem [{\citenamefont {Berkovits}\ and\ \citenamefont
  {Feng}(1990)}]{PhysRevLett.65.3120}%
  \BibitemOpen
  \bibfield  {author} {\bibinfo {author} {\bibfnamefont {R.}~\bibnamefont
  {Berkovits}}\ and\ \bibinfo {author} {\bibfnamefont {S.}~\bibnamefont
  {Feng}},\ }\href {https://doi.org/10.1103/PhysRevLett.65.3120} {\bibfield
  {journal} {\bibinfo  {journal} {Phys. Rev. Lett.}\ }\textbf {\bibinfo
  {volume} {65}},\ \bibinfo {pages} {3120} (\bibinfo {year}
  {1990})}\BibitemShut {NoStop}%
\bibitem [{\citenamefont {Berkovits}(1991)}]{PhysRevB.43.8638}%
  \BibitemOpen
  \bibfield  {author} {\bibinfo {author} {\bibfnamefont {R.}~\bibnamefont
  {Berkovits}},\ }\href {https://doi.org/10.1103/PhysRevB.43.8638} {\bibfield
  {journal} {\bibinfo  {journal} {Phys. Rev. B}\ }\textbf {\bibinfo {volume}
  {43}},\ \bibinfo {pages} {8638} (\bibinfo {year} {1991})}\BibitemShut
  {NoStop}%
\bibitem [{\citenamefont {den Outer}\ \emph {et~al.}(1993)\citenamefont {den
  Outer}, \citenamefont {Nieuwenhuizen},\ and\ \citenamefont
  {Lagendijk}}]{denOuter:93}%
  \BibitemOpen
  \bibfield  {author} {\bibinfo {author} {\bibfnamefont {P.~N.}\ \bibnamefont
  {den Outer}}, \bibinfo {author} {\bibfnamefont {T.~M.}\ \bibnamefont
  {Nieuwenhuizen}},\ and\ \bibinfo {author} {\bibfnamefont {A.}~\bibnamefont
  {Lagendijk}},\ }\href {https://doi.org/10.1364/JOSAA.10.001209} {\bibfield
  {journal} {\bibinfo  {journal} {J. Opt. Soc. Am. A}\ }\textbf {\bibinfo
  {volume} {10}},\ \bibinfo {pages} {1209} (\bibinfo {year}
  {1993})}\BibitemShut {NoStop}%
\bibitem [{\citenamefont {Berk}\ and\ \citenamefont
  {Foreman}(2021)}]{berk2021multiple}%
  \BibitemOpen
  \bibfield  {author} {\bibinfo {author} {\bibfnamefont {J.}~\bibnamefont
  {Berk}}\ and\ \bibinfo {author} {\bibfnamefont {M.~R.}\ \bibnamefont
  {Foreman}},\ }\href@noop {} {\bibfield  {journal} {\bibinfo  {journal}
  {arXiv:2105.02798}\ } (\bibinfo {year} {2021})}\BibitemShut {NoStop}%
\bibitem [{\citenamefont {Szameit}\ \emph {et~al.}(2012)\citenamefont
  {Szameit}, \citenamefont {Shechtman}, \citenamefont {Osherovich},
  \citenamefont {Bullkich}, \citenamefont {Sidorenko}, \citenamefont {Dana},
  \citenamefont {Steiner}, \citenamefont {Kley}, \citenamefont {Gazit},
  \citenamefont {Cohen-Hyams}, \citenamefont {Shoham}, \citenamefont
  {Zibulevsky}, \citenamefont {Yavneh}, \citenamefont {Eldar}, \citenamefont
  {Cohen},\ and\ \citenamefont {Segev}}]{szameit_sparsity-based_2012}%
  \BibitemOpen
  \bibfield  {author} {\bibinfo {author} {\bibfnamefont {A.}~\bibnamefont
  {Szameit}}, \bibinfo {author} {\bibfnamefont {Y.}~\bibnamefont {Shechtman}},
  \bibinfo {author} {\bibfnamefont {E.}~\bibnamefont {Osherovich}}, \bibinfo
  {author} {\bibfnamefont {E.}~\bibnamefont {Bullkich}}, \bibinfo {author}
  {\bibfnamefont {P.}~\bibnamefont {Sidorenko}}, \bibinfo {author}
  {\bibfnamefont {H.}~\bibnamefont {Dana}}, \bibinfo {author} {\bibfnamefont
  {S.}~\bibnamefont {Steiner}}, \bibinfo {author} {\bibfnamefont {E.~B.}\
  \bibnamefont {Kley}}, \bibinfo {author} {\bibfnamefont {S.}~\bibnamefont
  {Gazit}}, \bibinfo {author} {\bibfnamefont {T.}~\bibnamefont {Cohen-Hyams}},
  \bibinfo {author} {\bibfnamefont {S.}~\bibnamefont {Shoham}}, \bibinfo
  {author} {\bibfnamefont {M.}~\bibnamefont {Zibulevsky}}, \bibinfo {author}
  {\bibfnamefont {I.}~\bibnamefont {Yavneh}}, \bibinfo {author} {\bibfnamefont
  {Y.~C.}\ \bibnamefont {Eldar}}, \bibinfo {author} {\bibfnamefont
  {O.}~\bibnamefont {Cohen}},\ and\ \bibinfo {author} {\bibfnamefont
  {M.}~\bibnamefont {Segev}},\ }\href {https://doi.org/10.1038/nmat3289}
  {\bibfield  {journal} {\bibinfo  {journal} {Nature Materials}\ }\textbf
  {\bibinfo {volume} {11}},\ \bibinfo {pages} {455} (\bibinfo {year}
  {2012})}\BibitemShut {NoStop}%
\bibitem [{\citenamefont {Zhang}\ \emph {et~al.}(2016)\citenamefont {Zhang},
  \citenamefont {Godavarthi}, \citenamefont {Chaumet}, \citenamefont {Maire},
  \citenamefont {Giovannini}, \citenamefont {Talneau}, \citenamefont {Allain},
  \citenamefont {Belkebir},\ and\ \citenamefont
  {Sentenac}}]{zhang_far-field_2016}%
  \BibitemOpen
  \bibfield  {author} {\bibinfo {author} {\bibfnamefont {T.}~\bibnamefont
  {Zhang}}, \bibinfo {author} {\bibfnamefont {C.}~\bibnamefont {Godavarthi}},
  \bibinfo {author} {\bibfnamefont {P.~C.}\ \bibnamefont {Chaumet}}, \bibinfo
  {author} {\bibfnamefont {G.}~\bibnamefont {Maire}}, \bibinfo {author}
  {\bibfnamefont {H.}~\bibnamefont {Giovannini}}, \bibinfo {author}
  {\bibfnamefont {A.}~\bibnamefont {Talneau}}, \bibinfo {author} {\bibfnamefont
  {M.}~\bibnamefont {Allain}}, \bibinfo {author} {\bibfnamefont
  {K.}~\bibnamefont {Belkebir}},\ and\ \bibinfo {author} {\bibfnamefont
  {A.}~\bibnamefont {Sentenac}},\ }\href
  {https://doi.org/10.1364/OPTICA.3.000609} {\bibfield  {journal} {\bibinfo
  {journal} {Optica}\ }\textbf {\bibinfo {volume} {3}},\ \bibinfo {pages} {609}
  (\bibinfo {year} {2016})}\BibitemShut {NoStop}%
\bibitem [{\citenamefont {Ober}\ \emph {et~al.}(2004)\citenamefont {Ober},
  \citenamefont {Ram},\ and\ \citenamefont {Ward}}]{ober_localization_2004}%
  \BibitemOpen
  \bibfield  {author} {\bibinfo {author} {\bibfnamefont {R.~J.}\ \bibnamefont
  {Ober}}, \bibinfo {author} {\bibfnamefont {S.}~\bibnamefont {Ram}},\ and\
  \bibinfo {author} {\bibfnamefont {E.~S.}\ \bibnamefont {Ward}},\ }\href
  {https://doi.org/10.1016/S0006-3495(04)74193-4} {\bibfield  {journal}
  {\bibinfo  {journal} {Biophysical Journal}\ }\textbf {\bibinfo {volume}
  {86}},\ \bibinfo {pages} {1185} (\bibinfo {year} {2004})}\BibitemShut
  {NoStop}%
\bibitem [{\citenamefont {Mortensen}\ \emph {et~al.}(2010)\citenamefont
  {Mortensen}, \citenamefont {Churchman}, \citenamefont {Spudich},\ and\
  \citenamefont {Flyvbjerg}}]{mortensen_optimized_2010}%
  \BibitemOpen
  \bibfield  {author} {\bibinfo {author} {\bibfnamefont {K.~I.}\ \bibnamefont
  {Mortensen}}, \bibinfo {author} {\bibfnamefont {L.~S.}\ \bibnamefont
  {Churchman}}, \bibinfo {author} {\bibfnamefont {J.~A.}\ \bibnamefont
  {Spudich}},\ and\ \bibinfo {author} {\bibfnamefont {H.}~\bibnamefont
  {Flyvbjerg}},\ }\href {https://doi.org/10.1038/nmeth.1447} {\bibfield
  {journal} {\bibinfo  {journal} {Nature Methods}\ }\textbf {\bibinfo {volume}
  {7}},\ \bibinfo {pages} {377} (\bibinfo {year} {2010})}\BibitemShut {NoStop}%
\bibitem [{\citenamefont {Deschout}\ \emph {et~al.}(2014)\citenamefont
  {Deschout}, \citenamefont {Zanacchi}, \citenamefont {Mlodzianoski},
  \citenamefont {Diaspro}, \citenamefont {Bewersdorf}, \citenamefont {Hess},\
  and\ \citenamefont {Braeckmans}}]{deschout_precisely_2014}%
  \BibitemOpen
  \bibfield  {author} {\bibinfo {author} {\bibfnamefont {H.}~\bibnamefont
  {Deschout}}, \bibinfo {author} {\bibfnamefont {F.~C.}\ \bibnamefont
  {Zanacchi}}, \bibinfo {author} {\bibfnamefont {M.}~\bibnamefont
  {Mlodzianoski}}, \bibinfo {author} {\bibfnamefont {A.}~\bibnamefont
  {Diaspro}}, \bibinfo {author} {\bibfnamefont {J.}~\bibnamefont {Bewersdorf}},
  \bibinfo {author} {\bibfnamefont {S.~T.}\ \bibnamefont {Hess}},\ and\
  \bibinfo {author} {\bibfnamefont {K.}~\bibnamefont {Braeckmans}},\ }\href
  {https://doi.org/10.1038/nmeth.2843} {\bibfield  {journal} {\bibinfo
  {journal} {Nature Methods}\ }\textbf {\bibinfo {volume} {11}},\ \bibinfo
  {pages} {253} (\bibinfo {year} {2014})}\BibitemShut {NoStop}%
\bibitem [{\citenamefont {Backlund}\ \emph {et~al.}(2018)\citenamefont
  {Backlund}, \citenamefont {Shechtman},\ and\ \citenamefont
  {Walsworth}}]{backlund_fundamental_2018}%
  \BibitemOpen
  \bibfield  {author} {\bibinfo {author} {\bibfnamefont {M.~P.}\ \bibnamefont
  {Backlund}}, \bibinfo {author} {\bibfnamefont {Y.}~\bibnamefont
  {Shechtman}},\ and\ \bibinfo {author} {\bibfnamefont {R.~L.}\ \bibnamefont
  {Walsworth}},\ }\href {https://doi.org/10.1103/PhysRevLett.121.023904}
  {\bibfield  {journal} {\bibinfo  {journal} {Phys. Rev. Lett.}\ }\textbf
  {\bibinfo {volume} {121}},\ \bibinfo {pages} {023904} (\bibinfo {year}
  {2018})}\BibitemShut {NoStop}%
\bibitem [{\citenamefont {Ram}\ \emph {et~al.}(2006)\citenamefont {Ram},
  \citenamefont {Ward},\ and\ \citenamefont {Ober}}]{ram_beyond_2006}%
  \BibitemOpen
  \bibfield  {author} {\bibinfo {author} {\bibfnamefont {S.}~\bibnamefont
  {Ram}}, \bibinfo {author} {\bibfnamefont {E.~S.}\ \bibnamefont {Ward}},\ and\
  \bibinfo {author} {\bibfnamefont {R.~J.}\ \bibnamefont {Ober}},\ }\href
  {https://doi.org/10.1073/pnas.0508047103} {\bibfield  {journal} {\bibinfo
  {journal} {Proceedings of the National Academy of Sciences}\ }\textbf
  {\bibinfo {volume} {103}},\ \bibinfo {pages} {4457} (\bibinfo {year}
  {2006})}\BibitemShut {NoStop}%
\bibitem [{\citenamefont {Tsang}\ \emph {et~al.}(2016)\citenamefont {Tsang},
  \citenamefont {Nair},\ and\ \citenamefont {Lu}}]{tsang_quantum_2016}%
  \BibitemOpen
  \bibfield  {author} {\bibinfo {author} {\bibfnamefont {M.}~\bibnamefont
  {Tsang}}, \bibinfo {author} {\bibfnamefont {R.}~\bibnamefont {Nair}},\ and\
  \bibinfo {author} {\bibfnamefont {X.-M.}\ \bibnamefont {Lu}},\ }\href
  {https://doi.org/10.1103/PhysRevX.6.031033} {\bibfield  {journal} {\bibinfo
  {journal} {Physical Review X}\ }\textbf {\bibinfo {volume} {6}},\ \bibinfo
  {pages} {031033} (\bibinfo {year} {2016})}\BibitemShut {NoStop}%
\bibitem [{\citenamefont {Paúr}\ \emph {et~al.}(2016)\citenamefont {Paúr},
  \citenamefont {Stoklasa}, \citenamefont {Hradil}, \citenamefont
  {Sánchez-Soto},\ and\ \citenamefont {Rehacek}}]{paur_achieving_2016}%
  \BibitemOpen
  \bibfield  {author} {\bibinfo {author} {\bibfnamefont {M.}~\bibnamefont
  {Paúr}}, \bibinfo {author} {\bibfnamefont {B.}~\bibnamefont {Stoklasa}},
  \bibinfo {author} {\bibfnamefont {Z.}~\bibnamefont {Hradil}}, \bibinfo
  {author} {\bibfnamefont {L.~L.}\ \bibnamefont {Sánchez-Soto}},\ and\
  \bibinfo {author} {\bibfnamefont {J.}~\bibnamefont {Rehacek}},\ }\href
  {https://doi.org/10.1364/OPTICA.3.001144} {\bibfield  {journal} {\bibinfo
  {journal} {Optica}\ }\textbf {\bibinfo {volume} {3}},\ \bibinfo {pages}
  {1144} (\bibinfo {year} {2016})}\BibitemShut {NoStop}%
\bibitem [{\citenamefont {Zhou}\ \emph {et~al.}(2019)\citenamefont {Zhou},
  \citenamefont {Yang}, \citenamefont {Hassett}, \citenamefont {Rafsanjani},
  \citenamefont {Mirhosseini}, \citenamefont {Vamivakas}, \citenamefont
  {Jordan}, \citenamefont {Shi},\ and\ \citenamefont {Boyd}}]{Zhou:19}%
  \BibitemOpen
  \bibfield  {author} {\bibinfo {author} {\bibfnamefont {Y.}~\bibnamefont
  {Zhou}}, \bibinfo {author} {\bibfnamefont {J.}~\bibnamefont {Yang}}, \bibinfo
  {author} {\bibfnamefont {J.~D.}\ \bibnamefont {Hassett}}, \bibinfo {author}
  {\bibfnamefont {S.~M.~H.}\ \bibnamefont {Rafsanjani}}, \bibinfo {author}
  {\bibfnamefont {M.}~\bibnamefont {Mirhosseini}}, \bibinfo {author}
  {\bibfnamefont {A.~N.}\ \bibnamefont {Vamivakas}}, \bibinfo {author}
  {\bibfnamefont {A.~N.}\ \bibnamefont {Jordan}}, \bibinfo {author}
  {\bibfnamefont {Z.}~\bibnamefont {Shi}},\ and\ \bibinfo {author}
  {\bibfnamefont {R.~W.}\ \bibnamefont {Boyd}},\ }\href
  {https://doi.org/10.1364/OPTICA.6.000534} {\bibfield  {journal} {\bibinfo
  {journal} {Optica}\ }\textbf {\bibinfo {volume} {6}},\ \bibinfo {pages} {534}
  (\bibinfo {year} {2019})}\BibitemShut {NoStop}%
\bibitem [{\citenamefont {Sentenac}\ \emph {et~al.}(2007)\citenamefont
  {Sentenac}, \citenamefont {Guérin}, \citenamefont {Chaumet}, \citenamefont
  {Drsek}, \citenamefont {Giovannini}, \citenamefont {Bertaux},\ and\
  \citenamefont {Holschneider}}]{sentenac_influence_2007}%
  \BibitemOpen
  \bibfield  {author} {\bibinfo {author} {\bibfnamefont {A.}~\bibnamefont
  {Sentenac}}, \bibinfo {author} {\bibfnamefont {C.-A.}\ \bibnamefont
  {Guérin}}, \bibinfo {author} {\bibfnamefont {P.~C.}\ \bibnamefont
  {Chaumet}}, \bibinfo {author} {\bibfnamefont {F.}~\bibnamefont {Drsek}},
  \bibinfo {author} {\bibfnamefont {H.}~\bibnamefont {Giovannini}}, \bibinfo
  {author} {\bibfnamefont {N.}~\bibnamefont {Bertaux}},\ and\ \bibinfo {author}
  {\bibfnamefont {M.}~\bibnamefont {Holschneider}},\ }\href
  {https://doi.org/10.1364/OE.15.001340} {\bibfield  {journal} {\bibinfo
  {journal} {Optics Express}\ }\textbf {\bibinfo {volume} {15}},\ \bibinfo
  {pages} {1340} (\bibinfo {year} {2007})}\BibitemShut {NoStop}%
\bibitem [{\citenamefont {Bouchet}\ \emph {et~al.}(2020)\citenamefont
  {Bouchet}, \citenamefont {Carminati},\ and\ \citenamefont
  {Mosk}}]{bouchet_influence_2020}%
  \BibitemOpen
  \bibfield  {author} {\bibinfo {author} {\bibfnamefont {D.}~\bibnamefont
  {Bouchet}}, \bibinfo {author} {\bibfnamefont {R.}~\bibnamefont {Carminati}},\
  and\ \bibinfo {author} {\bibfnamefont {A.~P.}\ \bibnamefont {Mosk}},\ }\href
  {https://doi.org/10.1103/PhysRevLett.124.133903} {\bibfield  {journal}
  {\bibinfo  {journal} {Physical Review Letters}\ }\textbf {\bibinfo {volume}
  {124}},\ \bibinfo {pages} {133903} (\bibinfo {year} {2020})}\BibitemShut
  {NoStop}%
\bibitem [{\citenamefont {Bouchet}\ \emph {et~al.}(2021)\citenamefont
  {Bouchet}, \citenamefont {Rotter},\ and\ \citenamefont
  {Mosk}}]{bouchet_maximum_2021}%
  \BibitemOpen
  \bibfield  {author} {\bibinfo {author} {\bibfnamefont {D.}~\bibnamefont
  {Bouchet}}, \bibinfo {author} {\bibfnamefont {S.}~\bibnamefont {Rotter}},\
  and\ \bibinfo {author} {\bibfnamefont {A.~P.}\ \bibnamefont {Mosk}},\ }\href
  {https://doi.org/10.1038/s41567-020-01137-4} {\bibfield  {journal} {\bibinfo
  {journal} {Nature Physics}\ }\textbf {\bibinfo {volume} {17}},\ \bibinfo
  {pages} {564} (\bibinfo {year} {2021})}\BibitemShut {NoStop}%
\bibitem [{\citenamefont {Helstrom}(1969)}]{helstrom_quantum_1969}%
  \BibitemOpen
  \bibfield  {author} {\bibinfo {author} {\bibfnamefont {C.~W.}\ \bibnamefont
  {Helstrom}},\ }\href {https://doi.org/10.1007/BF01007479} {\bibfield
  {journal} {\bibinfo  {journal} {J. Stat. Phys.}\ }\textbf {\bibinfo {volume}
  {1}},\ \bibinfo {pages} {231} (\bibinfo {year} {1969})}\BibitemShut {NoStop}%
\bibitem [{\citenamefont {Barnes}(1998)}]{barnes_fluorescence_1998}%
  \BibitemOpen
  \bibfield  {author} {\bibinfo {author} {\bibfnamefont {W.~L.}\ \bibnamefont
  {Barnes}},\ }\href {https://doi.org/10.1080/09500349808230614} {\bibfield
  {journal} {\bibinfo  {journal} {J. Mod. Opt.}\ }\textbf {\bibinfo {volume}
  {45}},\ \bibinfo {pages} {661} (\bibinfo {year} {1998})}\BibitemShut
  {NoStop}%
\bibitem [{\citenamefont {Cazé}\ \emph {et~al.}(2013)\citenamefont {Cazé},
  \citenamefont {Pierrat},\ and\ \citenamefont
  {Carminati}}]{caze_spatial_2013}%
  \BibitemOpen
  \bibfield  {author} {\bibinfo {author} {\bibfnamefont {A.}~\bibnamefont
  {Cazé}}, \bibinfo {author} {\bibfnamefont {R.}~\bibnamefont {Pierrat}},\
  and\ \bibinfo {author} {\bibfnamefont {R.}~\bibnamefont {Carminati}},\ }\href
  {https://doi.org/10.1103/PhysRevLett.110.063903} {\bibfield  {journal}
  {\bibinfo  {journal} {Physical Review Letters}\ }\textbf {\bibinfo {volume}
  {110}},\ \bibinfo {pages} {043823} (\bibinfo {year} {2013})}\BibitemShut
  {NoStop}%
\bibitem [{\citenamefont {Braunstein}\ \emph {et~al.}(1996)\citenamefont
  {Braunstein}, \citenamefont {Caves},\ and\ \citenamefont
  {Milburn}}]{braunstein_generalized_1996}%
  \BibitemOpen
  \bibfield  {author} {\bibinfo {author} {\bibfnamefont {S.~L.}\ \bibnamefont
  {Braunstein}}, \bibinfo {author} {\bibfnamefont {C.~M.}\ \bibnamefont
  {Caves}},\ and\ \bibinfo {author} {\bibfnamefont {G.~J.}\ \bibnamefont
  {Milburn}},\ }\href {https://doi.org/10.1006/aphy.1996.0040} {\bibfield
  {journal} {\bibinfo  {journal} {Ann. Phys. (N. Y.)}\ }\textbf {\bibinfo
  {volume} {247}},\ \bibinfo {pages} {135} (\bibinfo {year}
  {1996})}\BibitemShut {NoStop}%
\bibitem [{\citenamefont {Trees}\ \emph {et~al.}(2013)\citenamefont {Trees},
  \citenamefont {Bell},\ and\ \citenamefont {Tian}}]{trees_detection_2013}%
  \BibitemOpen
  \bibfield  {author} {\bibinfo {author} {\bibfnamefont {H.~L.~V.}\
  \bibnamefont {Trees}}, \bibinfo {author} {\bibfnamefont {K.~L.}\ \bibnamefont
  {Bell}},\ and\ \bibinfo {author} {\bibfnamefont {Z.}~\bibnamefont {Tian}},\
  }\href@noop {} {\emph {\bibinfo {title} {Detection {Estimation} and
  {Modulation} {Theory}, {Part} {I}}}}\ (\bibinfo  {publisher} {Wiley},\
  \bibinfo {year} {2013})\BibitemShut {NoStop}%
\bibitem [{\citenamefont {Ambichl}\ \emph {et~al.}(2017)\citenamefont
  {Ambichl}, \citenamefont {Brandstötter}, \citenamefont {Böhm},
  \citenamefont {Kühmayer}, \citenamefont {Kuhl},\ and\ \citenamefont
  {Rotter}}]{ambichl_focusing_2017}%
  \BibitemOpen
  \bibfield  {author} {\bibinfo {author} {\bibfnamefont {P.}~\bibnamefont
  {Ambichl}}, \bibinfo {author} {\bibfnamefont {A.}~\bibnamefont
  {Brandstötter}}, \bibinfo {author} {\bibfnamefont {J.}~\bibnamefont
  {Böhm}}, \bibinfo {author} {\bibfnamefont {M.}~\bibnamefont {Kühmayer}},
  \bibinfo {author} {\bibfnamefont {U.}~\bibnamefont {Kuhl}},\ and\ \bibinfo
  {author} {\bibfnamefont {S.}~\bibnamefont {Rotter}},\ }\href
  {https://doi.org/10.1103/PhysRevLett.119.033903} {\bibfield  {journal}
  {\bibinfo  {journal} {Physical Review Letters}\ }\textbf {\bibinfo {volume}
  {119}},\ \bibinfo {pages} {033903} (\bibinfo {year} {2017})}\BibitemShut
  {NoStop}%
\bibitem [{\citenamefont {Horodynski}\ \emph {et~al.}(2020)\citenamefont
  {Horodynski}, \citenamefont {Kühmayer}, \citenamefont {Brandstötter},
  \citenamefont {Pichler}, \citenamefont {Fyodorov}, \citenamefont {Kuhl},\
  and\ \citenamefont {Rotter}}]{horodynski_optimal_2020}%
  \BibitemOpen
  \bibfield  {author} {\bibinfo {author} {\bibfnamefont {M.}~\bibnamefont
  {Horodynski}}, \bibinfo {author} {\bibfnamefont {M.}~\bibnamefont
  {Kühmayer}}, \bibinfo {author} {\bibfnamefont {A.}~\bibnamefont
  {Brandstötter}}, \bibinfo {author} {\bibfnamefont {K.}~\bibnamefont
  {Pichler}}, \bibinfo {author} {\bibfnamefont {Y.~V.}\ \bibnamefont
  {Fyodorov}}, \bibinfo {author} {\bibfnamefont {U.}~\bibnamefont {Kuhl}},\
  and\ \bibinfo {author} {\bibfnamefont {S.}~\bibnamefont {Rotter}},\ }\href
  {https://doi.org/10.1038/s41566-019-0550-z} {\bibfield  {journal} {\bibinfo
  {journal} {Nature Photonics}\ }\textbf {\bibinfo {volume} {14}},\ \bibinfo
  {pages} {149} (\bibinfo {year} {2020})}\BibitemShut {NoStop}%
\bibitem [{\citenamefont {Savo}\ \emph {et~al.}(2017)\citenamefont {Savo},
  \citenamefont {Pierrat}, \citenamefont {Najar}, \citenamefont {Carminati},
  \citenamefont {Rotter},\ and\ \citenamefont {Gigan}}]{savo_observation_2017}%
  \BibitemOpen
  \bibfield  {author} {\bibinfo {author} {\bibfnamefont {R.}~\bibnamefont
  {Savo}}, \bibinfo {author} {\bibfnamefont {R.}~\bibnamefont {Pierrat}},
  \bibinfo {author} {\bibfnamefont {U.}~\bibnamefont {Najar}}, \bibinfo
  {author} {\bibfnamefont {R.}~\bibnamefont {Carminati}}, \bibinfo {author}
  {\bibfnamefont {S.}~\bibnamefont {Rotter}},\ and\ \bibinfo {author}
  {\bibfnamefont {S.}~\bibnamefont {Gigan}},\ }\href
  {https://doi.org/10.1126/science.aan4054} {\bibfield  {journal} {\bibinfo
  {journal} {Science}\ }\textbf {\bibinfo {volume} {358}},\ \bibinfo {pages}
  {765} (\bibinfo {year} {2017})}\BibitemShut {NoStop}%
\bibitem [{\citenamefont {Bénichou}\ \emph {et~al.}(2005)\citenamefont
  {Bénichou}, \citenamefont {Coppey}, \citenamefont {Moreau}, \citenamefont
  {Suet},\ and\ \citenamefont {Voituriez}}]{benichou_averaged_2005}%
  \BibitemOpen
  \bibfield  {author} {\bibinfo {author} {\bibfnamefont {O.}~\bibnamefont
  {Bénichou}}, \bibinfo {author} {\bibfnamefont {M.}~\bibnamefont {Coppey}},
  \bibinfo {author} {\bibfnamefont {M.}~\bibnamefont {Moreau}}, \bibinfo
  {author} {\bibfnamefont {P.~H.}\ \bibnamefont {Suet}},\ and\ \bibinfo
  {author} {\bibfnamefont {R.}~\bibnamefont {Voituriez}},\ }\href
  {https://doi.org/10.1209/epl/i2005-10001-y} {\bibfield  {journal} {\bibinfo
  {journal} {Europhysics Letters (EPL)}\ }\textbf {\bibinfo {volume} {70}},\
  \bibinfo {pages} {42} (\bibinfo {year} {2005})}\BibitemShut {NoStop}%
\bibitem [{\citenamefont {Canaguier-Durand}\ \emph {et~al.}(2019)\citenamefont
  {Canaguier-Durand}, \citenamefont {Pierrat},\ and\ \citenamefont
  {Carminati}}]{canaguier-durand_cross_2019}%
  \BibitemOpen
  \bibfield  {author} {\bibinfo {author} {\bibfnamefont {A.}~\bibnamefont
  {Canaguier-Durand}}, \bibinfo {author} {\bibfnamefont {R.}~\bibnamefont
  {Pierrat}},\ and\ \bibinfo {author} {\bibfnamefont {R.}~\bibnamefont
  {Carminati}},\ }\href {https://doi.org/10.1103/PhysRevA.99.013835} {\bibfield
   {journal} {\bibinfo  {journal} {Physical Review A}\ }\textbf {\bibinfo
  {volume} {99}},\ \bibinfo {pages} {013835} (\bibinfo {year}
  {2019})}\BibitemShut {NoStop}%
\bibitem [{\citenamefont {Weyl}(1911)}]{weyl_ueber_1911}%
  \BibitemOpen
  \bibfield  {author} {\bibinfo {author} {\bibfnamefont {H.}~\bibnamefont
  {Weyl}},\ }\href {http://eudml.org/doc/58792} {\bibfield  {journal} {\bibinfo
   {journal} {Nachrichten von der Gesellschaft der Wissenschaften zu
  Göttingen, Mathematisch-Physikalische Klasse}\ }\textbf {\bibinfo {volume}
  {1911}},\ \bibinfo {pages} {110} (\bibinfo {year} {1911})}\BibitemShut
  {NoStop}%
\bibitem [{\citenamefont {Arendt}\ and\ \citenamefont
  {Schleich}(2009)}]{arendt_mathematical_2009}%
  \BibitemOpen
  \bibinfo {editor} {\bibfnamefont {W.}~\bibnamefont {Arendt}}\ and\ \bibinfo
  {editor} {\bibfnamefont {W.~P.}\ \bibnamefont {Schleich}},\ eds.,\ \href
  {https://doi.org/10.1002/9783527628025} {\emph {\bibinfo {title}
  {Mathematical {Analysis} of {Evolution}, {Information}, and {Complexity}}}},\
  \bibinfo {edition} {1st}\ ed.\ (\bibinfo  {publisher} {Wiley},\ \bibinfo
  {year} {2009})\BibitemShut {NoStop}%
\bibitem [{\citenamefont {van Tiggelen}\ and\ \citenamefont
  {Skipetrov}(2006)}]{van_tiggelen_fluctuations_2006}%
  \BibitemOpen
  \bibfield  {author} {\bibinfo {author} {\bibfnamefont {B.~A.}\ \bibnamefont
  {van Tiggelen}}\ and\ \bibinfo {author} {\bibfnamefont {S.~E.}\ \bibnamefont
  {Skipetrov}},\ }\href {https://doi.org/10.1103/PhysRevE.73.045601} {\bibfield
   {journal} {\bibinfo  {journal} {Physical Review E}\ }\textbf {\bibinfo
  {volume} {73}},\ \bibinfo {pages} {045601(R)} (\bibinfo {year}
  {2006})}\BibitemShut {NoStop}%
\bibitem [{\citenamefont {Cazé}\ \emph {et~al.}(2010)\citenamefont {Cazé},
  \citenamefont {Pierrat},\ and\ \citenamefont
  {Carminati}}]{caze_near-field_2010}%
  \BibitemOpen
  \bibfield  {author} {\bibinfo {author} {\bibfnamefont {A.}~\bibnamefont
  {Cazé}}, \bibinfo {author} {\bibfnamefont {R.}~\bibnamefont {Pierrat}},\
  and\ \bibinfo {author} {\bibfnamefont {R.}~\bibnamefont {Carminati}},\ }\href
  {https://doi.org/10.1103/PhysRevA.82.043823} {\bibfield  {journal} {\bibinfo
  {journal} {Physical Review A}\ }\textbf {\bibinfo {volume} {82}},\ \bibinfo
  {pages} {043823} (\bibinfo {year} {2010})}\BibitemShut {NoStop}%
\bibitem [{\citenamefont {Pierrat}\ \emph {et~al.}(2014)\citenamefont
  {Pierrat}, \citenamefont {Ambichl}, \citenamefont {Gigan}, \citenamefont
  {Haber}, \citenamefont {Carminati},\ and\ \citenamefont
  {Rotter}}]{pierrat_invariance_2014}%
  \BibitemOpen
  \bibfield  {author} {\bibinfo {author} {\bibfnamefont {R.}~\bibnamefont
  {Pierrat}}, \bibinfo {author} {\bibfnamefont {P.}~\bibnamefont {Ambichl}},
  \bibinfo {author} {\bibfnamefont {S.}~\bibnamefont {Gigan}}, \bibinfo
  {author} {\bibfnamefont {A.}~\bibnamefont {Haber}}, \bibinfo {author}
  {\bibfnamefont {R.}~\bibnamefont {Carminati}},\ and\ \bibinfo {author}
  {\bibfnamefont {S.}~\bibnamefont {Rotter}},\ }\href
  {https://doi.org/10.1073/pnas.1417725111} {\bibfield  {journal} {\bibinfo
  {journal} {Proceedings of the National Academy of Sciences}\ }\textbf
  {\bibinfo {volume} {111}},\ \bibinfo {pages} {17765} (\bibinfo {year}
  {2014})}\BibitemShut {NoStop}%
\bibitem [{\citenamefont {Davy}\ \emph {et~al.}(2021)\citenamefont {Davy},
  \citenamefont {Kühmayer}, \citenamefont {Gigan},\ and\ \citenamefont
  {Rotter}}]{davy_mean_2021}%
  \BibitemOpen
  \bibfield  {author} {\bibinfo {author} {\bibfnamefont {M.}~\bibnamefont
  {Davy}}, \bibinfo {author} {\bibfnamefont {M.}~\bibnamefont {Kühmayer}},
  \bibinfo {author} {\bibfnamefont {S.}~\bibnamefont {Gigan}},\ and\ \bibinfo
  {author} {\bibfnamefont {S.}~\bibnamefont {Rotter}},\ }\href
  {https://doi.org/10.1038/s42005-021-00585-5} {\bibfield  {journal} {\bibinfo
  {journal} {Communications Physics}\ }\textbf {\bibinfo {volume} {4}},\
  \bibinfo {pages} {85} (\bibinfo {year} {2021})}\BibitemShut {NoStop}%
\bibitem [{\citenamefont {Schöberl}(1997)}]{schoberl_netgen_1997}%
  \BibitemOpen
  \bibfield  {author} {\bibinfo {author} {\bibfnamefont {J.}~\bibnamefont
  {Schöberl}},\ }\href {https://doi.org/10.1007/s007910050004} {\bibfield
  {journal} {\bibinfo  {journal} {Computing and Visualization in Science}\
  }\textbf {\bibinfo {volume} {1}},\ \bibinfo {pages} {41} (\bibinfo {year}
  {1997})}\BibitemShut {NoStop}%
\bibitem [{\citenamefont {Schöberl}(2014)}]{schoberl_c11_2014}%
  \BibitemOpen
  \bibfield  {author} {\bibinfo {author} {\bibfnamefont {J.}~\bibnamefont
  {Schöberl}},\ }\href@noop {} {\emph {\bibinfo {title} {C++11
  {Implementation} of {Finite} {Elements} in {NGSolve}}}},\ \bibinfo {type}
  {{ASC} {Report}}\ (\bibinfo  {institution} {Institute for Analysis and
  Scientific Computing, Vienna University of Technology},\ \bibinfo {year}
  {2014})\BibitemShut {NoStop}%
\bibitem [{\citenamefont {Sapienza}\ \emph {et~al.}(2011)\citenamefont
  {Sapienza}, \citenamefont {Bondareff}, \citenamefont {Pierrat}, \citenamefont
  {Habert}, \citenamefont {Carminati},\ and\ \citenamefont {van
  Hulst}}]{sapienza_long-tail_2011}%
  \BibitemOpen
  \bibfield  {author} {\bibinfo {author} {\bibfnamefont {R.}~\bibnamefont
  {Sapienza}}, \bibinfo {author} {\bibfnamefont {P.}~\bibnamefont {Bondareff}},
  \bibinfo {author} {\bibfnamefont {R.}~\bibnamefont {Pierrat}}, \bibinfo
  {author} {\bibfnamefont {B.}~\bibnamefont {Habert}}, \bibinfo {author}
  {\bibfnamefont {R.}~\bibnamefont {Carminati}},\ and\ \bibinfo {author}
  {\bibfnamefont {N.~F.}\ \bibnamefont {van Hulst}},\ }\href
  {https://doi.org/10.1103/PhysRevLett.106.163902} {\bibfield  {journal}
  {\bibinfo  {journal} {Physical Review Letters}\ }\textbf {\bibinfo {volume}
  {106}},\ \bibinfo {pages} {163902} (\bibinfo {year} {2011})}\BibitemShut
  {NoStop}%
\bibitem [{\citenamefont {Mosk}\ \emph {et~al.}(2012)\citenamefont {Mosk},
  \citenamefont {Lagendijk}, \citenamefont {Lerosey},\ and\ \citenamefont
  {Fink}}]{mosk_controlling_2012}%
  \BibitemOpen
  \bibfield  {author} {\bibinfo {author} {\bibfnamefont {A.~P.}\ \bibnamefont
  {Mosk}}, \bibinfo {author} {\bibfnamefont {A.}~\bibnamefont {Lagendijk}},
  \bibinfo {author} {\bibfnamefont {G.}~\bibnamefont {Lerosey}},\ and\ \bibinfo
  {author} {\bibfnamefont {M.}~\bibnamefont {Fink}},\ }\href
  {https://doi.org/10.1038/nphoton.2012.88} {\bibfield  {journal} {\bibinfo
  {journal} {Nat. Photonics}\ }\textbf {\bibinfo {volume} {6}},\ \bibinfo
  {pages} {283} (\bibinfo {year} {2012})}\BibitemShut {NoStop}%
\bibitem [{\citenamefont {Rotter}\ and\ \citenamefont
  {Gigan}(2017)}]{rotter_light_2017}%
  \BibitemOpen
  \bibfield  {author} {\bibinfo {author} {\bibfnamefont {S.}~\bibnamefont
  {Rotter}}\ and\ \bibinfo {author} {\bibfnamefont {S.}~\bibnamefont {Gigan}},\
  }\href {https://doi.org/10.1103/RevModPhys.89.015005} {\bibfield  {journal}
  {\bibinfo  {journal} {Reviews of Modern Physics}\ }\textbf {\bibinfo {volume}
  {89}},\ \bibinfo {pages} {015005} (\bibinfo {year} {2017})}\BibitemShut
  {NoStop}%
\bibitem [{\citenamefont {Koenderink}\ \emph {et~al.}(2005)\citenamefont
  {Koenderink}, \citenamefont {Lagendijk},\ and\ \citenamefont
  {Vos}}]{koenderink_optical_2005}%
  \BibitemOpen
  \bibfield  {author} {\bibinfo {author} {\bibfnamefont {A.~F.}\ \bibnamefont
  {Koenderink}}, \bibinfo {author} {\bibfnamefont {A.}~\bibnamefont
  {Lagendijk}},\ and\ \bibinfo {author} {\bibfnamefont {W.~L.}\ \bibnamefont
  {Vos}},\ }\href {https://doi.org/10.1103/PhysRevB.72.153102} {\bibfield
  {journal} {\bibinfo  {journal} {Physical Review B}\ }\textbf {\bibinfo
  {volume} {72}},\ \bibinfo {pages} {153102} (\bibinfo {year}
  {2005})}\BibitemShut {NoStop}%
\bibitem [{\citenamefont {Barthelemy}\ \emph {et~al.}(2008)\citenamefont
  {Barthelemy}, \citenamefont {Bertolotti},\ and\ \citenamefont
  {Wiersma}}]{barthelemy_levy_2008}%
  \BibitemOpen
  \bibfield  {author} {\bibinfo {author} {\bibfnamefont {P.}~\bibnamefont
  {Barthelemy}}, \bibinfo {author} {\bibfnamefont {J.}~\bibnamefont
  {Bertolotti}},\ and\ \bibinfo {author} {\bibfnamefont {D.~S.}\ \bibnamefont
  {Wiersma}},\ }\href {https://doi.org/10.1038/nature06948} {\bibfield
  {journal} {\bibinfo  {journal} {Nature}\ }\textbf {\bibinfo {volume} {453}},\
  \bibinfo {pages} {495} (\bibinfo {year} {2008})}\BibitemShut {NoStop}%
\bibitem [{\citenamefont {Torquato}\ and\ \citenamefont
  {Stillinger}(2003)}]{torquato_local_2003}%
  \BibitemOpen
  \bibfield  {author} {\bibinfo {author} {\bibfnamefont {S.}~\bibnamefont
  {Torquato}}\ and\ \bibinfo {author} {\bibfnamefont {F.~H.}\ \bibnamefont
  {Stillinger}},\ }\href {https://doi.org/10.1103/PhysRevE.68.041113}
  {\bibfield  {journal} {\bibinfo  {journal} {Phys. Rev. E}\ }\textbf {\bibinfo
  {volume} {68}},\ \bibinfo {pages} {041113} (\bibinfo {year}
  {2003})}\BibitemShut {NoStop}%
\bibitem [{\citenamefont {Huang}\ \emph {et~al.}(2020)\citenamefont {Huang},
  \citenamefont {Tian}, \citenamefont {Gopar}, \citenamefont {Fang},\ and\
  \citenamefont {Genack}}]{huang_invariance_2020}%
  \BibitemOpen
  \bibfield  {author} {\bibinfo {author} {\bibfnamefont {Y.}~\bibnamefont
  {Huang}}, \bibinfo {author} {\bibfnamefont {C.}~\bibnamefont {Tian}},
  \bibinfo {author} {\bibfnamefont {V.~A.}\ \bibnamefont {Gopar}}, \bibinfo
  {author} {\bibfnamefont {P.}~\bibnamefont {Fang}},\ and\ \bibinfo {author}
  {\bibfnamefont {A.~Z.}\ \bibnamefont {Genack}},\ }\href
  {https://doi.org/10.1103/PhysRevLett.124.057401} {\bibfield  {journal}
  {\bibinfo  {journal} {Physical Review Letters}\ }\textbf {\bibinfo {volume}
  {124}},\ \bibinfo {pages} {057401} (\bibinfo {year} {2020})}\BibitemShut
  {NoStop}%
\end{thebibliography}

\end{document}


\title{Invariance property of the Fisher information in scattering media:\\ supplementary material}

\author{Michael Horodynski}
\affiliation{Institute for Theoretical Physics, Vienna University of Technology (TU Wien), 1040, Vienna, Austria}
\author{Dorian Bouchet}
\affiliation{Université Grenoble Alpes, CNRS, LIPhy, Grenoble, France}
\author{Stefan Rotter}
\affiliation{Institute for Theoretical Physics, Vienna University of Technology (TU Wien), 1040, Vienna, Austria}

\maketitle

\section{Derivation of Eq.~(3)}\label{ap:apa}

Here we provide the detailed derivation of the main result presented in Eq.~(3) of the main text. Our starting point is the decomposition of the generalized Wigner-Smith (GWS) operator into the Green's function $G$, the coupling matrix to the leads $V$ and the perturbation $\Delta H$ \cite{horodynski_optimal_2020},
\begin{align}
	Q_\theta = \frac{2}{\Delta \theta} V^\dagger G^\dagger \Delta H G V\,,
\end{align}
where $\Delta \theta$ quantifies the (infinitesimal) change in the system. In order to find the average Fisher information we need an expression for the trace of the Fisher information operator, which, in turn, is proportional to the trace of the squared GWS operator, i.e., 
\begin{align}
	\tr F_\theta & = \frac{4}{\Delta \theta^2} \tr \left[ V^\dagger G^\dagger \Delta H G V  V^\dagger G^\dagger \Delta H G V \right], \nonumber \\
	& = \frac{4}{\Delta \theta^2} \tr \left\{ \left[\imag \left(G\right) \Delta H\right]^2 \right\}\,, \label{eq:eqa2}
\end{align}
where we have used the identity $ G V V^\dagger G^\dagger = - \imag G $ which is derived from $SS^\dagger = \mathbbm{1}$ to arrive at the second line of Eq.~\eqref{eq:eqa2}. 

We now turn our attention to a specific perturbation, namely the shift of a sub-wavelength particle which we characterize by $\Delta H = k^2 (\varepsilon-1) [\df{\mathbf{r} - \mathbf{r}_T- \Delta \mathbf{r} } - \df{\mathbf{r}-\mathbf{r}_T} ]$, where $\mathbf{r}_T$ is the position of the target and $\Delta \mathbf{r}$ is an infinitesimal displacement. Putting this $\Delta H$ into Eq.~\eqref{eq:eqa2} we get:
\begin{align}
	\tr F_x = & \frac{4 k^4 \left(\varepsilon-1\right)^2}{\Delta x^2 } \left[\imag{G\left(\mathbf{r}_T,\mathbf{r}_T;k^2\right)}^2 \right. \nonumber\\
	& - 2 \imag{G\left(\mathbf{r}_T,\mathbf{r}_T + \Delta \mathbf{r};k^2\right)}^2  \nonumber\\
	& \left. + \imag{G\left(\mathbf{r}+\Delta \mathbf{r} ,\mathbf{r} + \Delta \mathbf{r};k^2\right)}^2\right]\,. \label{eq:eqa3}
\end{align}
In this equation we now identify the Green's function $G$ with the so-called cross density of states (CDOS) \cite{caze_spatial_2013,canaguier-durand_cross_2019} defined as 
\begin{align}
	\rho_{12}\left(k\right) = -\frac{2k}{\pi} \imag \widetilde{G}\left(\mathbf{r}_1, \mathbf{r}_2;k^2\right),
\end{align}
where $\widetilde{G} = \varepsilon^{1/2} G \varepsilon^{1/2}$ denotes the Green's function suitably defined to calculate the LDOS \cite{de_vries_point_1998}. Plugging the CDOS into Eq.~\eqref{eq:eqa3} we then get
\begin{align}
	\tr F_x  = & \frac{2 k^2 \left( \varepsilon-1\right)^2 \pi^2}{\varepsilon^2 \Delta x^2} \left(\rho_{11}^2 - \rho_{12}^2 \right), \label{eq:eqa5}\\
	\avg{\tr F_x } = & \frac{k^4 \left(\varepsilon-1\right)^2 \pi^2}{\varepsilon^2} \avg{ \rho_{11}^2 }, \label{eq:eqa6}
\end{align}
where we used the observation that for short distances the CDOS follows on average a simple law, irrespective of the surrounding scattering environment \cite{canaguier-durand_cross_2019} to arrive at Eq.~\eqref{eq:eqa6},
\begin{align}
	\avg{\rho_{12}^2} = J_0 \left(kr\right)^2\avg{\rho_{11} \rho_{22}} \approx \left(1- k^2\Delta x^2/2 \right)\avg{\rho_{11} \rho_{22}},
\end{align}
where $J_0$ is the zeroth-order Bessel function of the first kind and we identify $\rho_{11} \equiv \rho(\mathbf{r}_T,k)$.\\\indent
In the case of measuring the dielectric constant, the perturbation is given by $\Delta H = k^2 \Delta \varepsilon \df{\mathbf{r} - \mathbf{r}_T}$. Employing the same reasoning as the one presented above in the case of position estimations, we obtain
\begin{equation}
    \avg{\tr F_\varepsilon} = \frac{k^2\pi^2}{\varepsilon^2} \avg{\rho_{11}^2}\,.
\end{equation}
This implies that the same invariance property as for $F_x$ also exists for measurements on the dielectric constant (apart from a different prefactor).

\section{Detailed analysis of LDOS scaling}\label{ap:apd}
Here we provide the detailed argument, why the Fisher information scales quadratically with the LDOS in the unitary case, while it scales linearly for non-unitary scattering matrices. We start by considering the derivative of the scattering matrix with respect to some arbitrary parameter:
\begin{equation}
    \partial_\theta S = -2i V^\dagger G \Delta H G V/\Delta \theta,
\end{equation}
This gives us the following expression for the Fisher information operator:
\begin{align}
    F_\theta & = \frac{4}{\Delta\theta^2} V^\dagger G^\dagger \Delta H G^\dagger V V^\dagger G \Delta H G V \\
    & = - \frac{4}{\Delta\theta^2} V^\dagger G^\dagger \Delta H \imag(G) \Delta H G V \label{eq:eqd11}\\
    & = \frac{2\pi}{k\Delta\theta^2} V^\dagger G^\dagger \Delta H \varepsilon^{-1/2} \rho \varepsilon^{-1/2} \Delta H G V, \label{eq:eqd12}
\end{align}
where we use $G^\dagger V V^\dagger G = - \imag(G)$, which follows from $S^\dagger S = \mathbbm{1}$ to arrive at Eq.~\eqref{eq:eqd11}. We reach Eq.~\eqref{eq:eqd12} by first transforming $G$ to $\widetilde{G}$ and then converting it to the CDOS. In the case of a point-like scatterer and $\theta=\varepsilon$, the perturbation is $\Delta H = k^2 \Delta \varepsilon\df{\mathbf{r}-\mathbf{r}_0}$, which results in
\begin{align}\label{eq:eqb5}
    \mathcal{I}(\varepsilon, \ket{u}) = \frac{2\pi k^3}{\varepsilon} |\psi_u(\mathbf{r}_0)|^2\rho(\mathbf{r}_0),
\end{align}
for the Fisher information of an arbitrary incidence wavefront described by $\ket{u}$, with $\psi_u$ being the resulting wave function. This shows that the scaling $\mathcal{I} \propto \rho^2$ comes from two contributions: The intensity at the target \cite{bouchet_influence_2020} and the unitarity of $S$. Both contribute a linear scaling of the QFI with the LDOS and together they result in $\mathcal{I} \propto \rho^2$. In the case of a non-unitary $S$ matrix only a linear scaling of the QFI with the LDOS can therefore be observed \cite{bouchet_influence_2020}. \\\indent
The preceding result also readily solves another subtle point. The eigenvalues of $F_\theta$ fulfill the following identity in the unitary case , which follows from $F_\theta=Q_\theta^2$: 
\begin{equation}
    \Lambda = \braket{\psi_u | \Delta H | \psi_u}^2/(4\Delta\theta^2),
\end{equation}
showing that eigenvalues scale quadratically with the intensity at the target while scaling linearly with the input intensity (which is $\propto \braket{u|u}$). Since the intensity at the target is proportional to the input intensity, this seems contradictory. Equation~\eqref{eq:eqb5} expresses the fact that different processes contribute to the intensity at the target position: the first contribution comes from the LDOS at the target position, while the second contribution is connected to the input intensity.

\section{Speckle correlations}\label{ap:apb}
\begin{figure}[t!]
	\centering
	\includegraphics[width=\columnwidth]{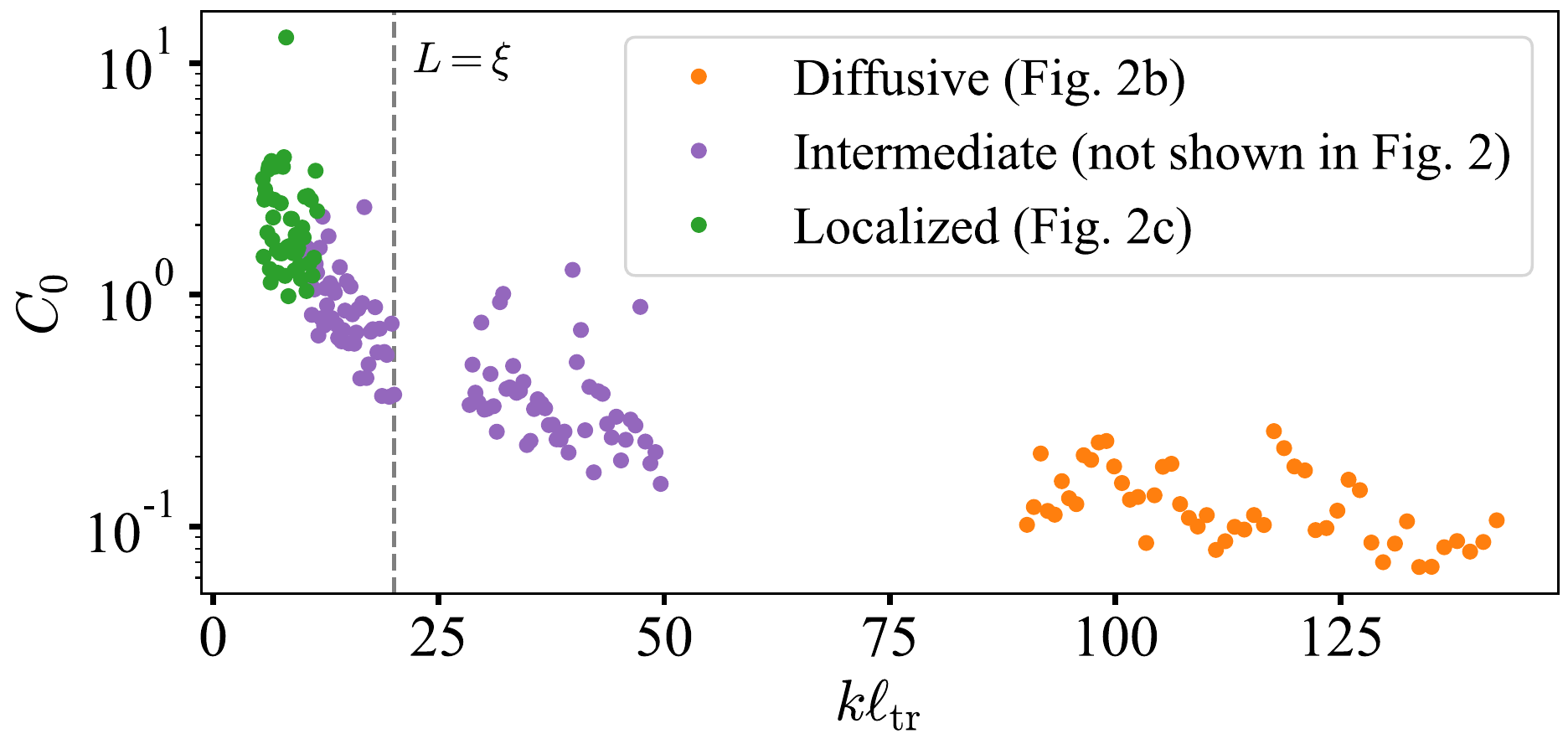}
	\caption{$C_0$–speckle correlation as a function of the normalised transport mean free path $k\ell_\mathrm{tr}$. Orange and green dots are obtained in the regimes considered in Fig.~2b and Fig.~2c of the manuscript, respectively. Purple dots are obtained from additional numerical simulations performed at the onset of the localized regime. The sharp increase of $C_0$ for small $k\ell_\mathrm{tr}$ explains the increase of the average quantum Fisher information observed in the localized regime (see Fig.~2c of the manuscript). Each dot is obtained after averaging over 100 random configurations. The vertical dashed grey line marks the onset of Anderson localization ($\xi=L$).}
	\label{fig:fig2}
\end{figure}

A central aspect of Eq.~(4) in the main manuscript is that the average quantum Fisher information depends on the $C_0$–speckle correlation. We thus numerically studied the dependence of $C_0 = \var [\rho(\mathbf{r}_T,k)]/\langle \rho(\mathbf{r}_T,k) \rangle^2$ upon the normalized transport mean free path $k\ell_\mathrm{tr}$. In the diffusive regime (for $k\ell_\mathrm{tr}$ of the order of $10^2$), the $C_0$–speckle correlation is negligible ($C_0 \simeq 10^{-1}$), as shown in Fig.~\ref{fig:fig2}. With decreasing $k\ell_\mathrm{tr}$, the value of $C_0$ starts to increase. Thus, in this regime, the average quantum Fisher information increases with the scattering strength of the environment, as predicted from Eq.~(4) and observed in Fig.~2c.

\section{Connection to QCRB}\label{ap:apc}
The QCRB is log-normal distributed, $\Sigma_\theta \sim \mathrm{Lognorm}(\mu, \sigma^2)$ \cite{bouchet_influence_2020}. This means that its inverse, which is the Fisher information, is also log-normal distributed with $\mu \rightarrow - \mu$, i.e., $\mathcal{I}_\theta \sim \mathrm{Lognorm}(-\mu, \sigma^2)$. This implies that the average localization precision given by the QCRB is related to the Fisher information by
\begin{equation}
    \avg{ \Sigma_\theta^\mathrm{avg} } = e^{-2\mu} \langle \mathcal{I}_\theta^\mathrm{avg} \rangle,
\end{equation}
where $e^{-2\mu} = 2^4 \avg{\tr^2 F_\theta}/\avg{\mathcal{I}_\theta^\mathrm{avg}}^4N^2$. In the case $\theta=x$  we get
\begin{equation}
    e^{-2\mu} = \frac{2^{6}N^2\left( \avg{\rho_{11}^4} -2 \avg{\rho_{11}^2 \rho_{12}^2} + \avg{\rho_{12}^4} \right) }{k^{20}R^{16}(\varepsilon-1)^4\pi^4 (1+C_0)^4 \Delta x^4 },
\end{equation}
while for $\theta=\varepsilon$ we find
\begin{equation}
    e^{-2\mu} = \frac{2^4 N^2\avg{\rho(\mathbf{r}_T,k)^4}}{k^{12}\varepsilon^4R^{16}\pi^4(1+C_0)^4}.
\end{equation}

\bibliographystyle{apsrev4-2}